\documentclass[aps,prb,reprint,superscriptaddress]{revtex4-2}
\usepackage{orcidlink}
\usepackage{amsmath,amssymb,bm}
\usepackage{graphicx}
\usepackage{physics}
\usepackage{hyperref}
\usepackage{color}
\usepackage{subcaption}

\begin{document}

\title{Localization Transition in Kinetically Deformed one-dimensional Aubry-Andr\'e Model}

\author{Arpita Goswami\,\orcidlink{0009-0003-2241-7034}}
\affiliation{Department of Physics, Indian Institute of Technology Tirupati, India, 517619}

\begin{abstract}
We propose a $q$-deformation in the single-particle kinetic energy and investigate how it modifies the localization in the one-dimensional Aubry-Andr\'e (AA) model. We construct a Hermitian $q$-deformed kinetic operator as a nonlinear function of the lattice translation operator, preserving the uniform lattice and recovering the conventional AA Hamiltonian continuously in the undeformed limit $q\to1$. The deformation generates an infinite set of correlated odd-range kinetic processes, controlled by a single parameter q, rather than phenomenologically involving long-range hopping. Under the dual transformation, this long-range hopping appears as higher harmonics of the dual quasiperiodic potential, providing a controlled route for breaking the exact self-duality of the AA model, with $q$ as the control parameter, consequently modifying the localization structure for $q\neq1$. In contrast to the conventional AA model, where all eigenstates localize simultaneously at $\lambda_c=2$, the deformed model exhibits a fraction of delocalized states even beyond $\lambda_c =2$. An intermediate regime also emerges in the $q-\lambda$ plane where localized and extended eigenstates coexist across the spectrum. We also propose a possible experimental realization of the hierarchy produced in the $q$-deformed kinetic setting in a periodically driven AA model. 

\end{abstract}

\maketitle

\section{Introduction}

Localization of quantum particles in quasiperiodic systems has been a central subject in condensed-matter physics since the introduction of the Aubry-Andr\'e (AA) model~\cite{Aubrey_1}. The one-dimensional AA model provides a paradigmatic example in which localization arises solely from deterministic quasiperiodicity, without the need for random disorder. In its standard form, the model possesses an exact self-duality between real and momentum space, leading to an energy-independent localization transition at $\lambda_c=2t$, where $t$ is the nearest-neighbor hopping amplitude and $\lambda$ denotes the strength of the quasiperiodic potential. This special correspondence between the kinetic and potential parts makes the AA model an exceptional reference point for understanding localization in one-dimensional quasiperiodic systems. It also raises a natural question: how is the localization structure modified when the conventional nearest-neighbor kinetic dispersion is replaced by a generalized one?

The special role of self-duality has motivated a broad class of generalized quasiperiodic models in which the hopping, the quasiperiodic potential, or both are modified. These extensions have revealed a rich variety of phenomena beyond the conventional AA scenario, including mobility edges, multifractal and critical states, intermediate localization regimes, and reentrant localization transitions~\cite{Mosaic_exact_ME_2, Exact_ME_1, exact_ME_5, Mosaic_ME_6, NH_exact_ME_4, PT_exact_ME_3, power_hop_ME_7, gen_AA, long_range_hopp_exp, long_range_kitaev_AAH, Long_range_AAH, goswami2025svql, goswami2026, subsystem_localization, Chatterjee_2023}. In most such constructions, the departure from the AA limit is introduced through spatial modulation of the quasiperiodic potential~\cite{gen_AA, Shallow_QP, sv_sds_1, sv_sds_2, thouless, goswami2025svql}, modified or independently controlled hopping amplitudes~\cite{QP_4, coupled_1d_ref_2, coupled_1d_sds, coupled_ref_3, NNN_AA, off_diag_2, off_diag_dis_also_sv}. In particular, long-range kinetic processes have been shown to generate nontrivial localization structures, including regimes with energy-dependent localization properties. Long-range quasiperiodic models can also be obtained from space-fractional quantum mechanics~\cite{Chatterjee_2023}, where a fractional spatial derivative leads to power-law kinetic processes. The present construction follows a distinct route: the lattice translation generator itself is deformed through a $q$-number prescription, producing an infinite correlated odd-range hopping hierarchy rather than a power-law hopping profile.

We employ $q$-deformed algebraic constructions as a framework for implementing such a controlled deformation. In particular, $q$-deformed quantum algebras employ the $q$-number as:
\begin{equation}
[x]_q=\frac{q^x-q^{-x}}{q-q^{-1}},
\label{eq:q_number}
\end{equation}
which continuously reduces to the ordinary number $x$ in the undeformed limit $q\to 1$~\cite{KANIADAKIS1997227}. Writing $q=e^\eta$, Eq.~\eqref{eq:q_number} can be expressed as:
\begin{equation}
[x]_q=\frac{\sinh(\eta x)}{\sinh\eta}.
\label{eq:q_number_hyperbolic}
\end{equation}
Such $q$-deformed algebraic structures have been employed in a variety of quantum-mechanical settings, including deformed oscillator and many-body systems, quantum gases, integrable models, and generalized quantum formalisms~\cite{free_electron, q_deformed_stat, q_deformed_harmonic, q_bose_gas, free_quon_gas, q_oscillator, q_boson_D_dimension, q_defromed_commutation, q_quantum_gas, Anyon_Fields, He4, quantum_chain, nuclear1, nuclear2, Conformal_theory, Quantum_group}. Motivated by this established algebraic structure, we use the $q$-number as a controlled deformation function for the lattice kinetic generator, and we then examine the fate of Anderson localization in the one-dimensional AA model under this implementation.

Now, we impose some requirements for the deformation to satisfy. First, it should remain Hermitian so that the resulting single-particle problem defines a conventional quantum Hamiltonian. Second, it should preserve the uniform underlying lattice rather than introducing an additional spatial inhomogeneity. Third, it should possess a controlled undeformed limit in which the conventional nearest-neighbor AA kinetic energy is recovered exactly. At the same time, the deformation should generate genuinely new kinetic processes rather than merely renormalizing the nearest-neighbor hopping. We realize these requirements using the unitary lattice translation operator. Specifically, we replace the conventional kinetic operator by a nonlinear function of the Hermitian combination $X=T+T^\dagger$. The resulting $q$-dependent kinetic energy continuously approaches the standard nearest-neighbor form as $q\to1$, while for $q\neq1$ it generates a correlated hierarchy of longer-range hopping processes. 

From an experimental perspective, the kinetic deformation considered here can be viewed as an effective description of a lattice in which single-particle tunneling extends beyond nearest neighbors. Such long-range tunneling processes can be engineered in Floquet quantum systems~\cite{Floquet_singleparticle1, Floquet_singleparticle2, Floquet_manybody_1, floquet_manybody2}, including ultracold atoms in optical lattices~\cite{ultracold_1, ultracold_2, ultracold_3, ultracold_4, ultracold_5, ultracold_6}, photonic waveguide arrays~\cite{expt_1, expt_2, expt_3, expt_4}, and programmable quantum circuit platforms~\cite{sc_qubit_1, sc_qubit_2}, where couplings between spatially separated sites can be controlled. In the present construction, however, the longer-range hopping amplitudes are not introduced independently; rather, they form a correlated hierarchy fixed by the single deformation parameter \(q\). Thus, the \(q\)-deformed kinetic operator provides a compact theoretical parametrization of a class of experimentally engineerable long-range tunneling structures.

We find that this kinetic deformation leads to qualitatively different localization behavior for $q\neq1$. In contrast to the conventional AA model, where the localization transition occurs simultaneously across the spectrum at $\lambda_c=2$, the deformed model exhibits a strongly energy-dependent localization structure. In addition, an intermediate regime emerges in which localized and extended eigenstates coexist across different portions of the spectrum, resulting in a finite fraction of delocalized states. Energy-resolved diagnostics further reveal that the boundary separating localized and extended states depends on energy, demonstrating that deformation of the kinetic part can qualitatively reorganize the eigenstate structure without modifying the underlying quasi-periodic potential.

\par
The remainder of the paper is organized as follows. In Sec.~\ref{sec2}, we introduce the conventional Aubry-Andr\'e model, and in Sec.~\ref{sec3} we formulate the $q$-deformed kinetic operator using the symmetric $q$-number prescription. We derive its real-space representation, identify the hierarchy of deformation-induced long-range hopping processes, and establish the undeformed limit $q\rightarrow1$. We also analyze the corresponding single-particle dispersion and its representation in momentum space, and examine the consequences of the deformation under the dual transformation, including the resulting breakdown of the standard AA self-duality. Sec.~\ref{sec_numeric} is devoted to the localization properties of the deformed model. We introduce the IPR and NPR diagnostics and investigate the evolution of the eigenstates as functions of deformation ($q$) and quasiperiodic potential strength ($\lambda$). Energy-resolved localization properties and finite-size behavior are analyzed to characterize the intermediate regime and the energy-dependent localization transition. We propose a possible physical realization of this deformation in a periodically driven system in sec.~\ref{floqt_realization}. In sec.~\ref{conc}, we summarize our main results and discuss their implications for kinetically deformed quasiperiodic systems and possible extensions of the present framework. Additional technical details on the $q$-number construction and finite-size convergence are provided in the Appendices.

\begin{figure}
    \centering
    \includegraphics[width=0.75\linewidth]{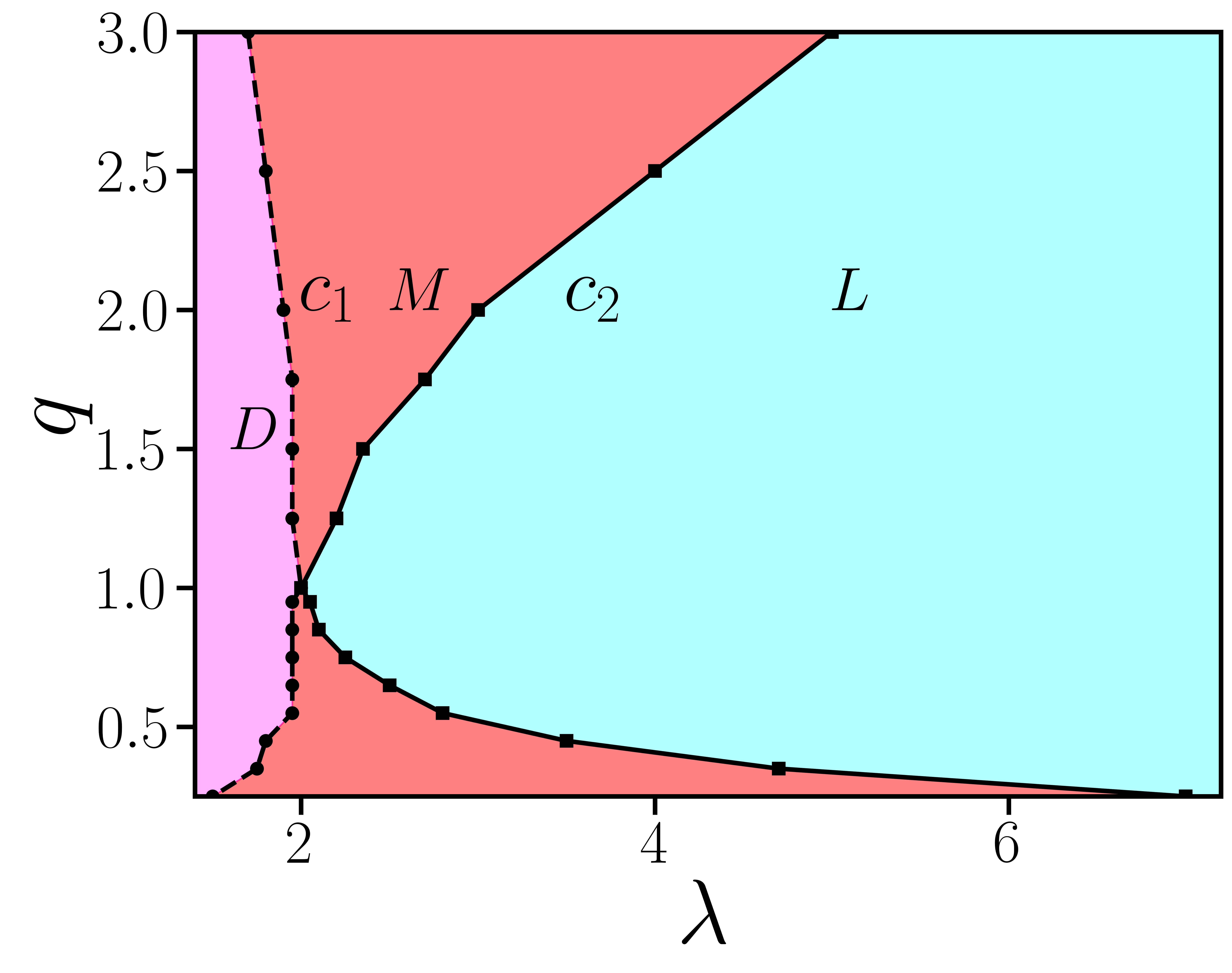}
    \caption{Numerically determined localization map to the q-generalized AA model. $D=$ delocalized phase, $M=$ mixed (intermediate) phase, and $L=$ Localized phase. The $c_1$ and $c_2$ lines correspond to the $D$-$M$ and $M$-$L$ crossover lines in the phase diagram.}
    \label{phase_diag}
\end{figure}

\begin{figure*}[t]
    \centering
    \begin{subfigure}[t]{0.45\linewidth}
        \includegraphics[width=\linewidth]{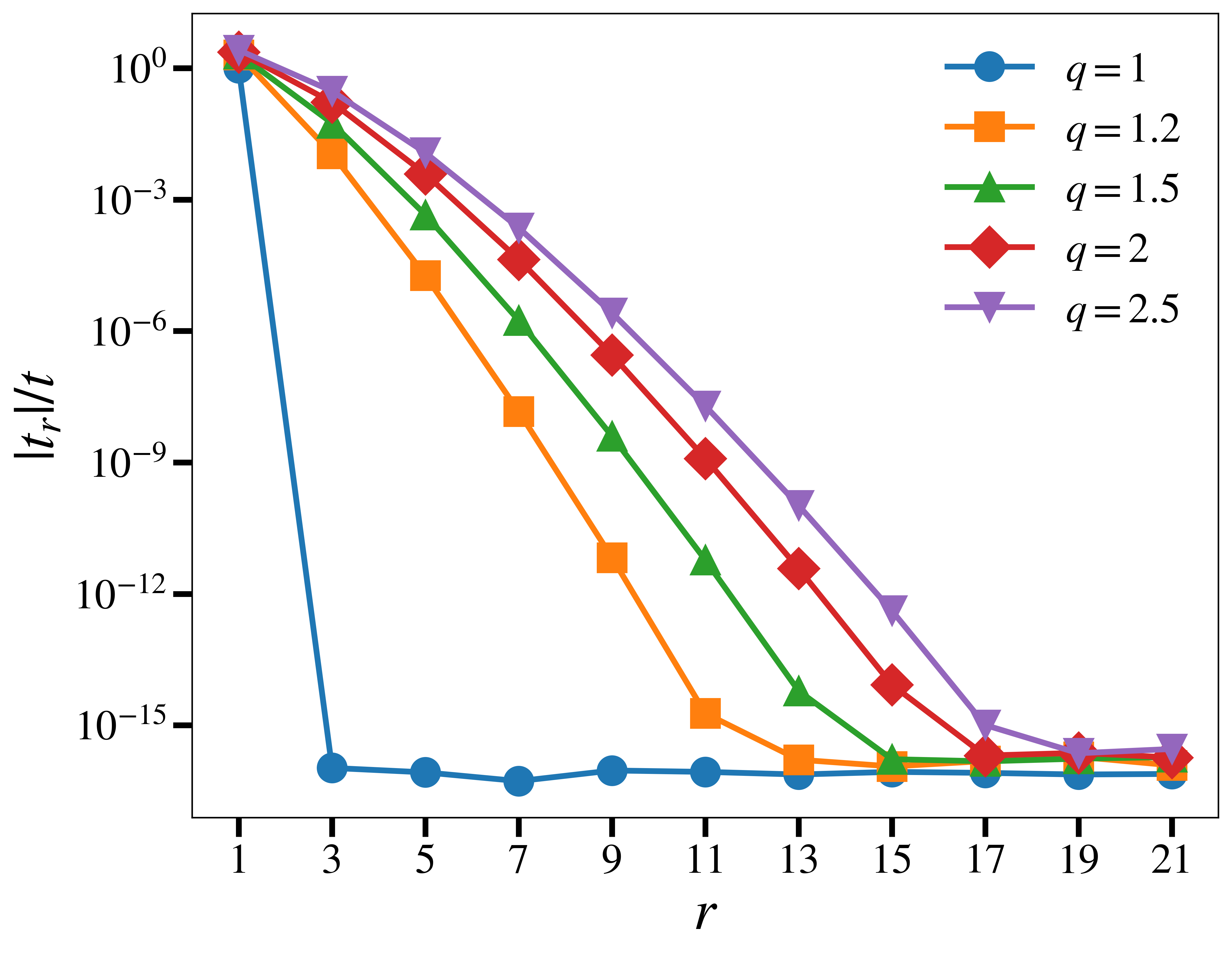}
        \caption{$|t_r/t_1|$ vs. r}
    \end{subfigure}
    \hspace{0.05\linewidth}
    \begin{subfigure}[t]{0.45\linewidth}
        \includegraphics[width=\linewidth]{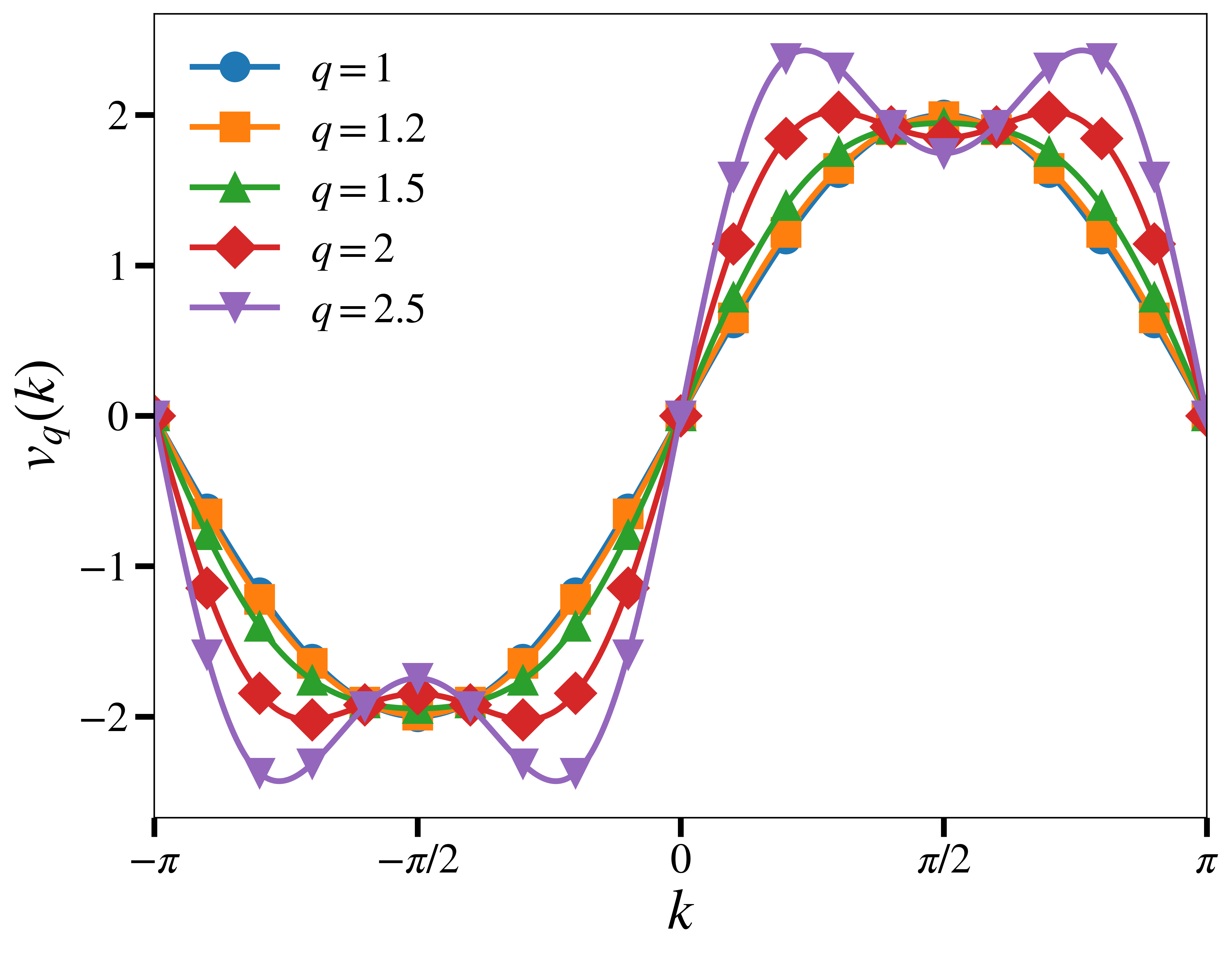}
        \caption{$v_q(k)$ vs. $k$}
    \end{subfigure}
    \caption{(a) Normalized odd-range hopping amplitudes $|t_r/t_1|$ generated by the $q$-deformed kinetic operator. The higher range processes are not independently tunable but are correlated through the single deformation parameter $q$.
    (b) Corresponding group velocity $v_q(k)=\partial\varepsilon_q(k)/\partial k$. The deformation progressively modifies the velocity's momentum dependence, shifting it away from the sinusoidal form of the conventional nearest-neighbor model. These two representations provide complementary views of the same kinetic deformation.}
    \label{fig:kinetic_deformation}
\end{figure*}

\begin{figure}
    \centering
    \includegraphics[width=0.5\textwidth]{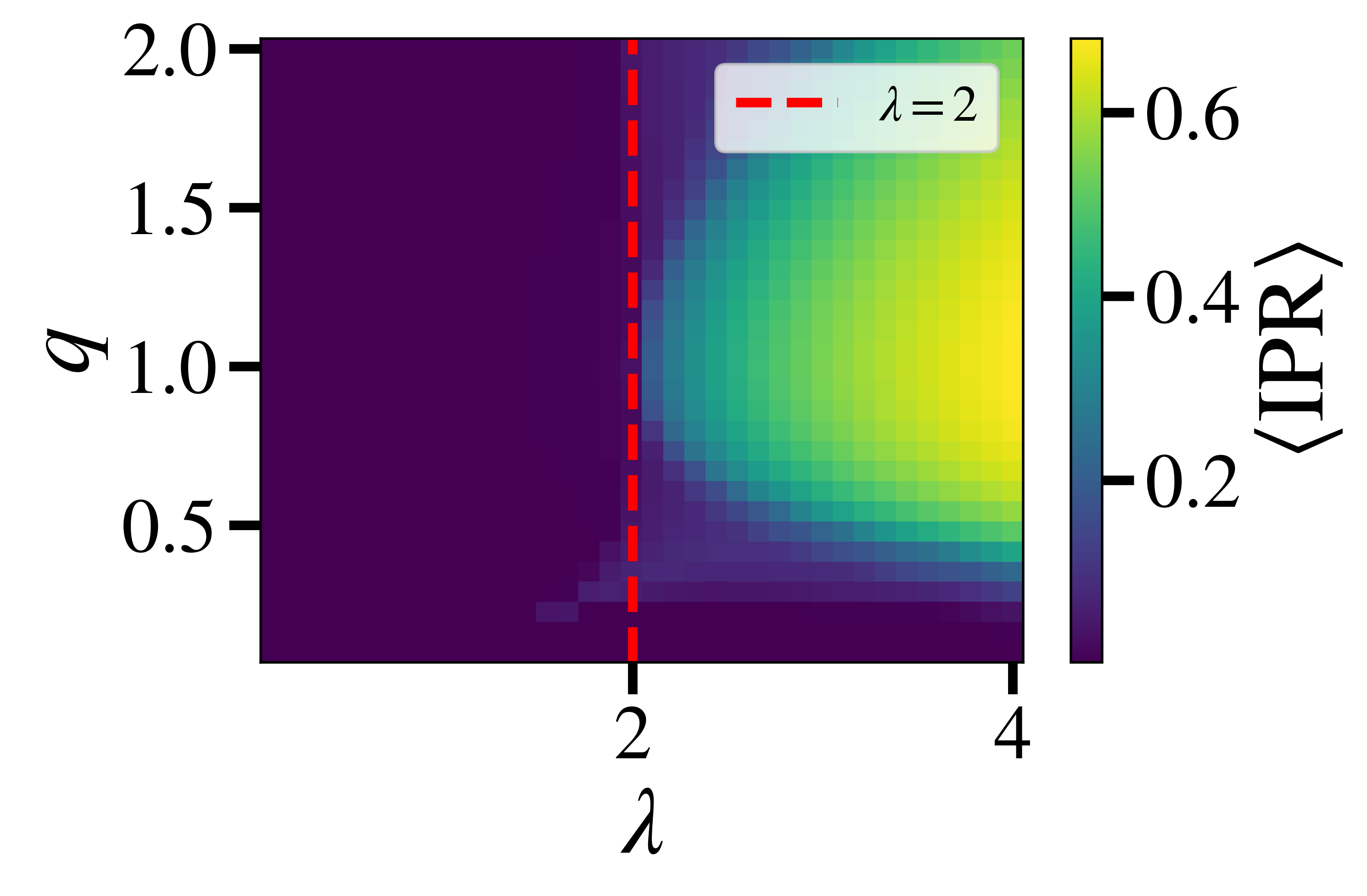}
    \caption{Average IPR as a function of $q$ and $\lambda.$ The red dashed line corresponds to $\lambda=2$. The $q$ is varied from $0.1$ to $2$ and $\lambda$ is varied from $0$ to $4$. $q=1$ gives back the standard AA limit. Calculations are done for $N = 1000$.}
    \label{fig:contour_ipr_q_lambda}
\end{figure}

\begin{figure*}[t]
    \centering
    \begin{subfigure}[t]{0.45\linewidth}
        \includegraphics[width=\linewidth]{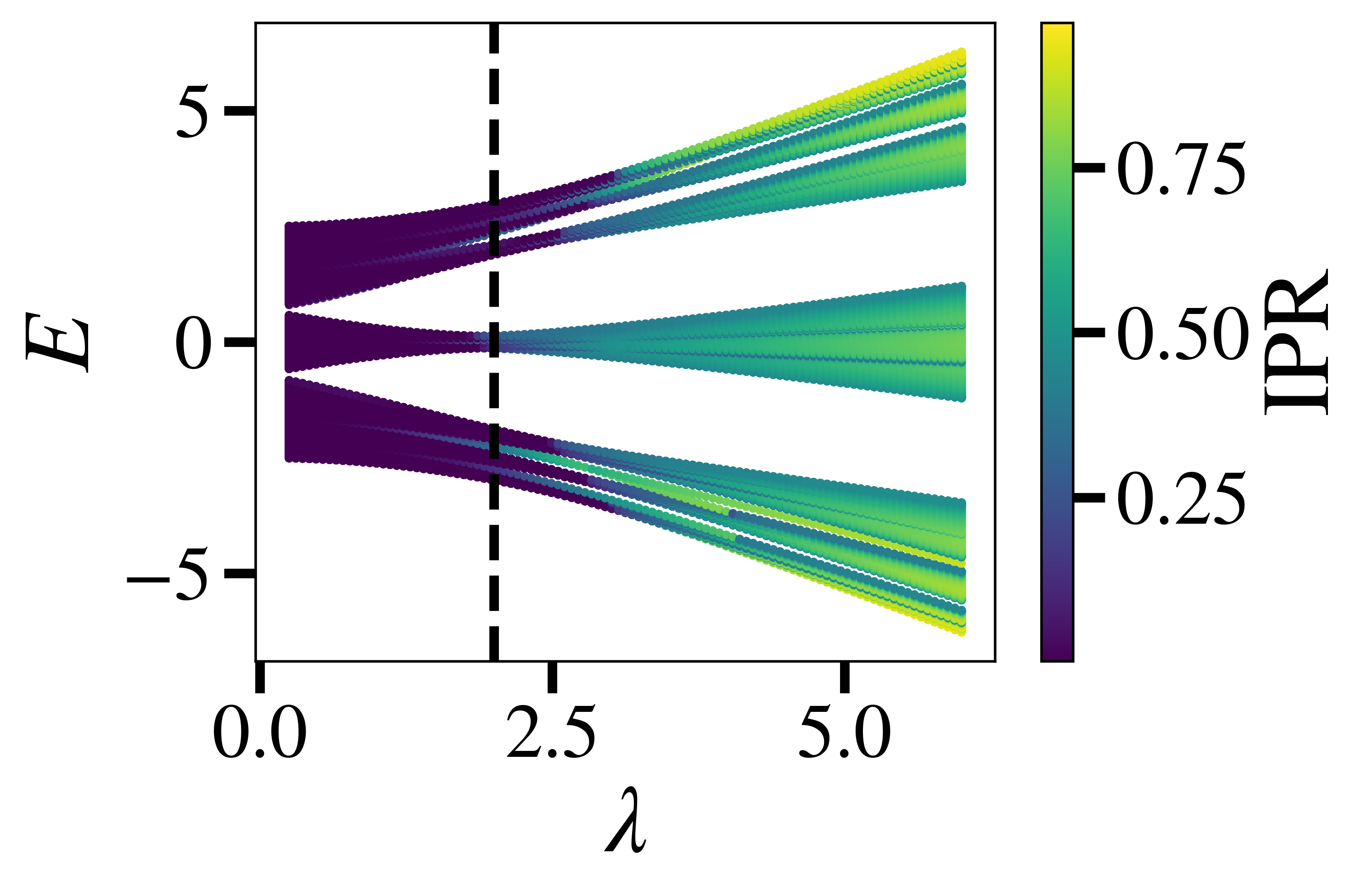}
        \caption{IPR as a function of $E$ and $\lambda$ for $q=0.5$.}
    \end{subfigure}
    \hspace{0.05\linewidth}
    \begin{subfigure}[t]{0.45\linewidth}
        \includegraphics[width=\linewidth]{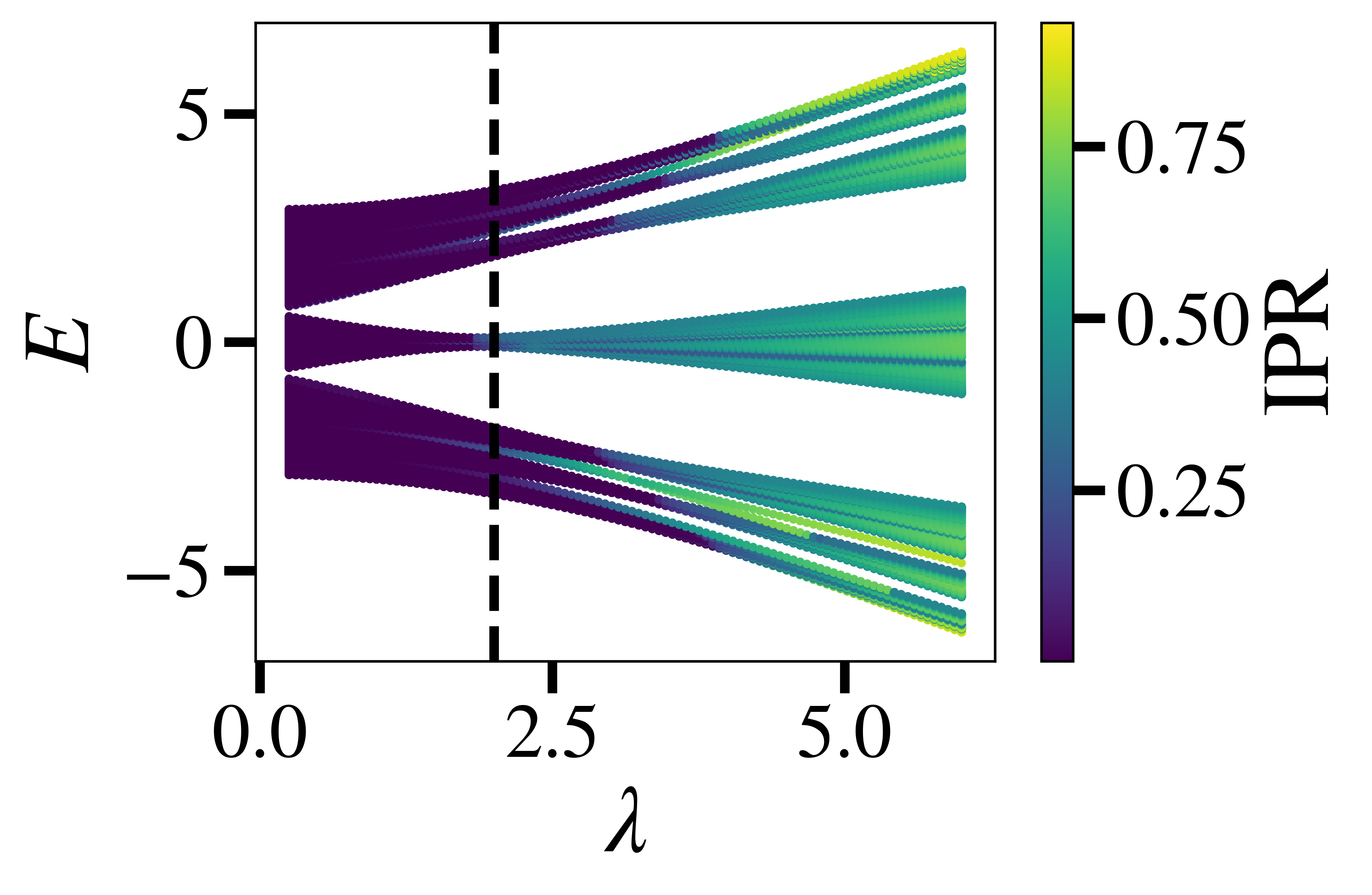}
        \caption{IPR as a function of $E$ and $\lambda$ for $q=2.5$.}
    \end{subfigure}
    \caption{The contour plot of IPR of each state as a function of energy ($E$) and quasi-periodic potential strength for (a) $q=0.5$ and (b) $q=2.5$ for system size $N=1000$. }
    \label{fig:fig2}
\end{figure*}

\begin{figure*}[t]
    \centering
    \begin{subfigure}[t]{0.45\linewidth}
        \includegraphics[width=\linewidth]{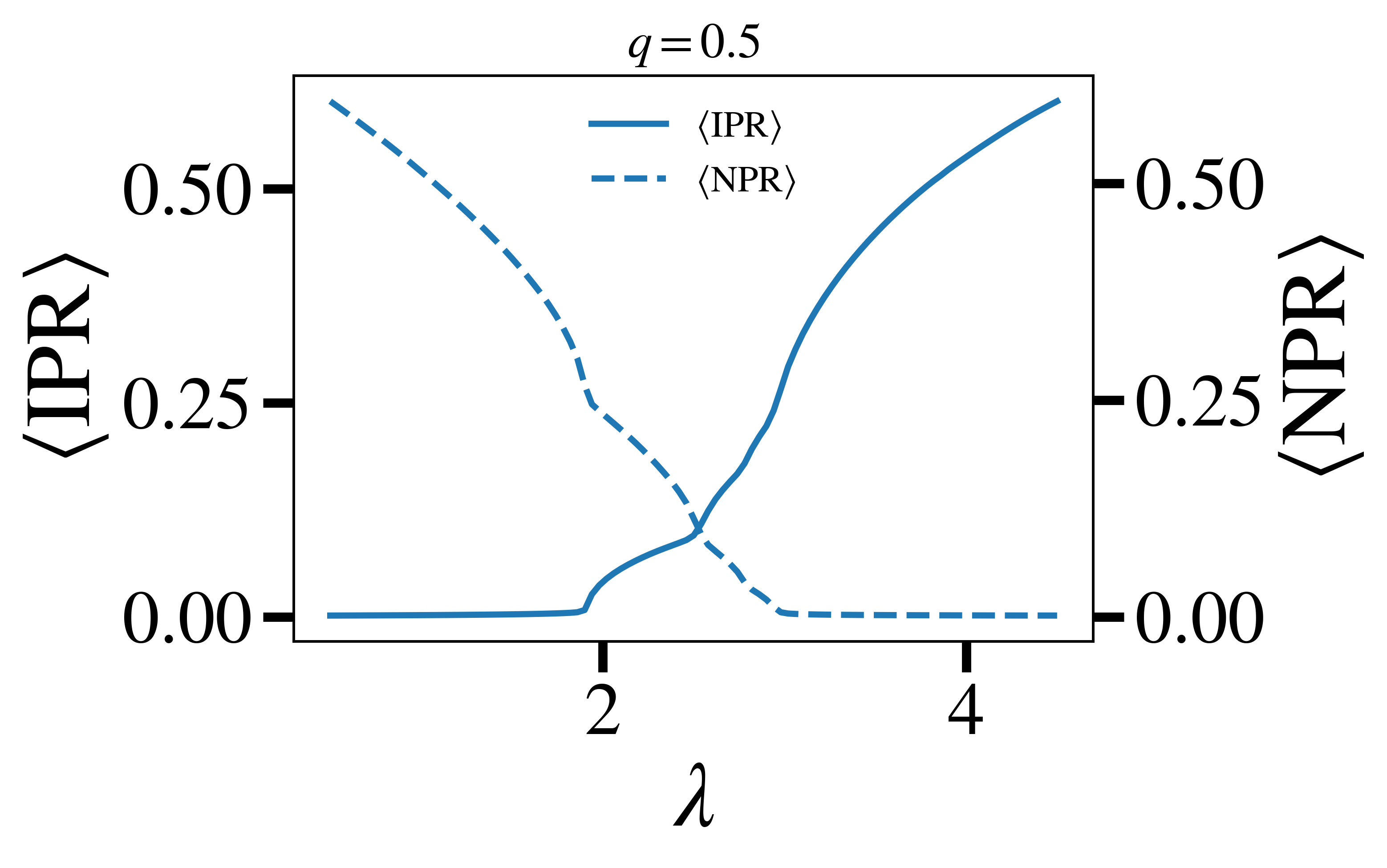}
        \caption{$\langle IPR \rangle$, $\langle NPR \rangle$ as a function of $\lambda$ at $q=0.5$.}
    \end{subfigure}
    \hspace{0.05\linewidth}
    \begin{subfigure}[t]{0.45\linewidth}
        \includegraphics[width=\linewidth]{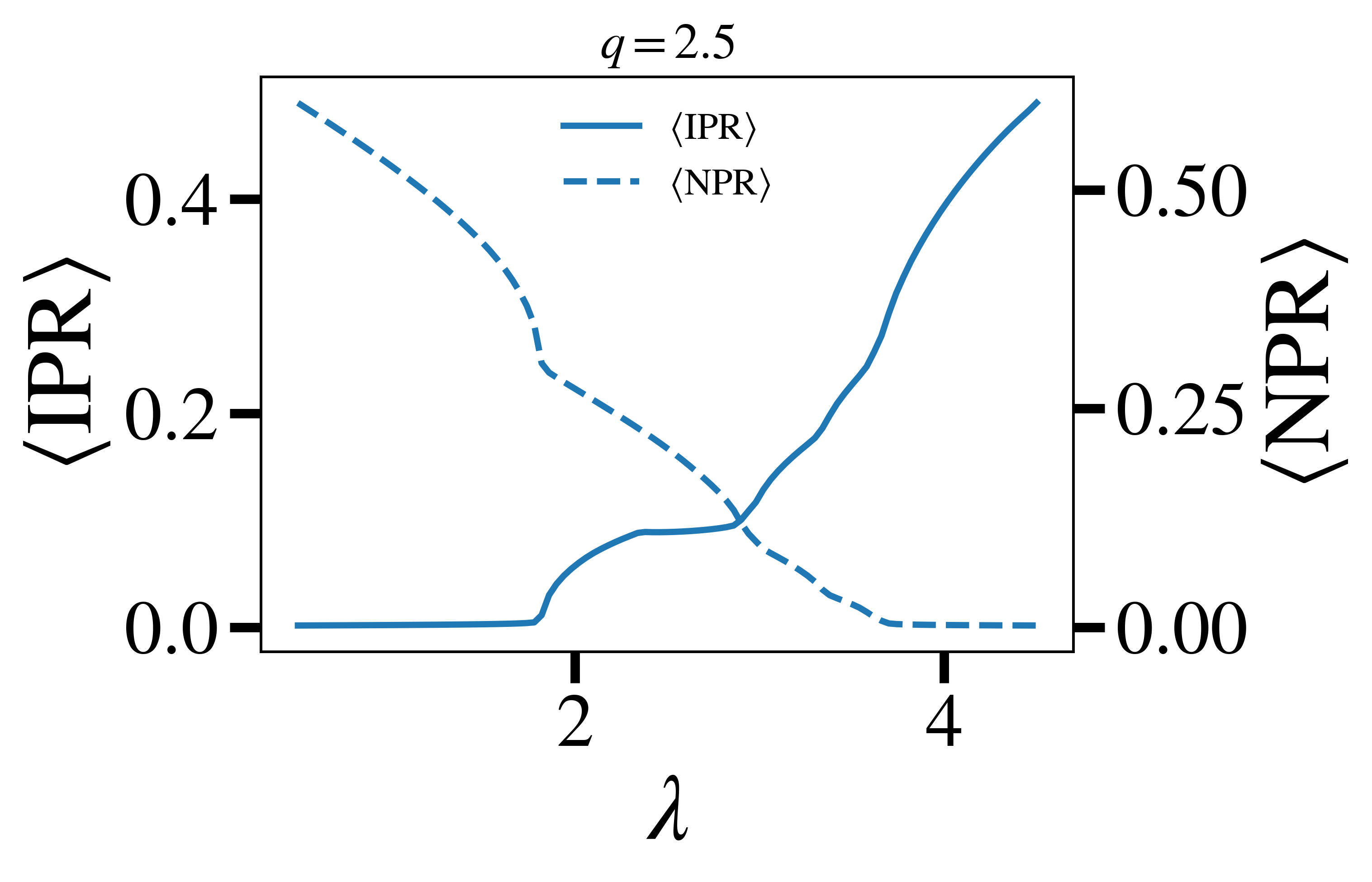}
        \caption{$\langle IPR \rangle$, $\langle NPR \rangle$ as a function of $\lambda$ at $q=2.5$.}
    \end{subfigure}
    \caption{ $\langle IPR \rangle$, $\langle NPR \rangle$ as a function of $\lambda$ at (a) $q=0.5$ and (b) $q=2.5$ respectively for system size $N=1000$.}
    \label{fig:fig3}
\end{figure*}

\section{The Aubry-Andr\'e model and lattice translation algebra}
\label{sec2}

The standard one-dimensional AA Hamiltonian is given by
\begin{equation}
\begin{split}
H_{\rm AA} = -t\sum_n \left(c_{n+1}^{\dagger}c_n+c_n^\dagger c_{n+1}\right) &\\
+ \lambda\sum_n \cos(2\pi\alpha n+\phi)c_n^\dagger c_n ,
\label{eq:AA}
\end{split}
\end{equation}
where $\alpha=\frac{\sqrt 5 -1}{2}$ is an irrational number and $\phi$ is the global phase factor and $t$ is nearest neighbor hopping amplitude and $\lambda$ is the strength of quasi-periodic potential. We fix $t=1$ for our calculation. In the single particle basis $\ket{n}$, the lattice translation operator $T$ can be defined as follows:
\begin{equation}
T\ket{n}=\ket{n+1}, \qquad T^\dagger\ket{n}=\ket{n-1}.
\end{equation}
The lattice translation operators satisfy
\begin{equation}
T^\dagger T=TT^\dagger=1.
\label{eq:unitarity}
\end{equation}

The lattice position operator $N$ satisfies the following eigenvalue equation
\begin{equation}
N\ket{n}=n\ket{n},
\end{equation}
Also, the lattice translational operator $T$ and the lattice position operator obey the following commutator relation
\begin{equation}
[N,T]=T, \qquad [N,T^\dagger]=-T^\dagger.
\label{eq:translationalgebra}
\end{equation}

The kinetic part of Eq.~(\ref{eq:AA}) can therefore be written as
\begin{equation}
H_{\rm kin} = -t(T+T^\dagger).
\label{eq:kinetic}
\end{equation}
As in momentum space, the translational operator is diagonal, so for a momentum eigenstate $\ket{k}$, the eigenvalue equation is given by
\begin{equation}
T\ket{k}=e^{ik}\ket{k},
\end{equation}
and consequently
\begin{equation}
H_{\rm kin}\ket{k} = -2t\cos k\ket{k}.
\end{equation}
Thus, the ordinary AA kinetic dispersion is
\begin{equation}
\epsilon(k)=-2t\cos k.
\label{eq:AAdispersion}
\end{equation}

\section{Kinetic deformation}
\label{sec3}

The deformation introduced below is motivated by the broader framework of q-calculus~\cite{q_derivative} and q-deformed physical systems~\cite{GOSWAMI2026131989, Goswami_2025}, where introducing a $ q-$deformation reveals richer phenomena and a more generalized framework. Rather than modifying the quasiperiodic potential or introducing an externally prescribed long-range hopping profile, we ask whether a \(q\)-dependent deformation can instead be incorporated directly into the lattice translational operator. For our implementation, we consider a one-dimensional Aubry-Andr\'e model. We incorporate this idea through the \(q\)-number construction. For a dimensionless operator \(X\), we define

\[[X]_q=\frac{q^X-q^{-X}}{q-q^{-1}}
=\frac{\sinh(\eta X)}{\sinh\eta},
\qquad q=e^\eta .\]

The ordinary operator is recovered continuously in the limit \(q\to1\), since \([X]_q\to X\). Choosing \(X=T+T^\dagger\), where \(T\) is the lattice translation operator, therefore provides a Hermitian \(q\)-deformation of the nearest-neighbor kinetic operator while preserving the uniform underlying lattice. We introduce the $q$-deformed kinetic operator as follows:
\begin{equation}
H_{\rm kin}^{(q)}=-t\,\frac{\sinh(\eta X)}{\sinh\eta},
\qquad q=e^\eta .
\label{eq:qkinetic}
\end{equation}

The operator $X$ is Hermitian, and hence $H_{\rm kin}^{(q)}$ is Hermitian for real $\eta$. The $q\to1$ limit corresponds to $\eta\to0$. For small $\eta$, using the Taylor series expansion we get
\begin{equation}
\frac{\sinh(\eta X)}{\sinh\eta}= X+\frac{\eta^2}{6}(X^3-X)+ \mathcal O(\eta^4),
\label{eq:qexpansion}
\end{equation}
Substituting in Eq.~\ref{eq:qkinetic}, we get
\begin{equation}
H_{\rm kin}^{(q)} = -tX - \frac{t\eta^2}{6}(X^3-X)
+\mathcal O(\eta^4).
\label{eq:kinexpansion}
\end{equation}
Since
$X=T+T^\dagger$ and $T$ and $T^\dagger$ commute,
\[X^3=T^3+3T+3T^\dagger+T^{\dagger3}.\]
Consequently,
\begin{equation}
X^3-X=T^3+2T+2T^\dagger+T^{\dagger3}.
\end{equation}

The kinetic operator, therefore, becomes
\begin{align} \label{eq:kinreal}
H_{\rm kin}^{(q)}  =&
-t\left(1+\frac{\eta^2}{3}\right) (T+T^\dagger)
\nonumber\\
&- \frac{t\eta^2}{6}  (T^3+T^{\dagger3})
+\mathcal O(\eta^4).
\end{align}

Thus, the deformation generates long-range hopping while preserving Hermiticity and translational invariance.

\begin{figure*}[t]
    \centering
    \begin{subfigure}[t]{0.45\linewidth}
        \includegraphics[width=\linewidth]{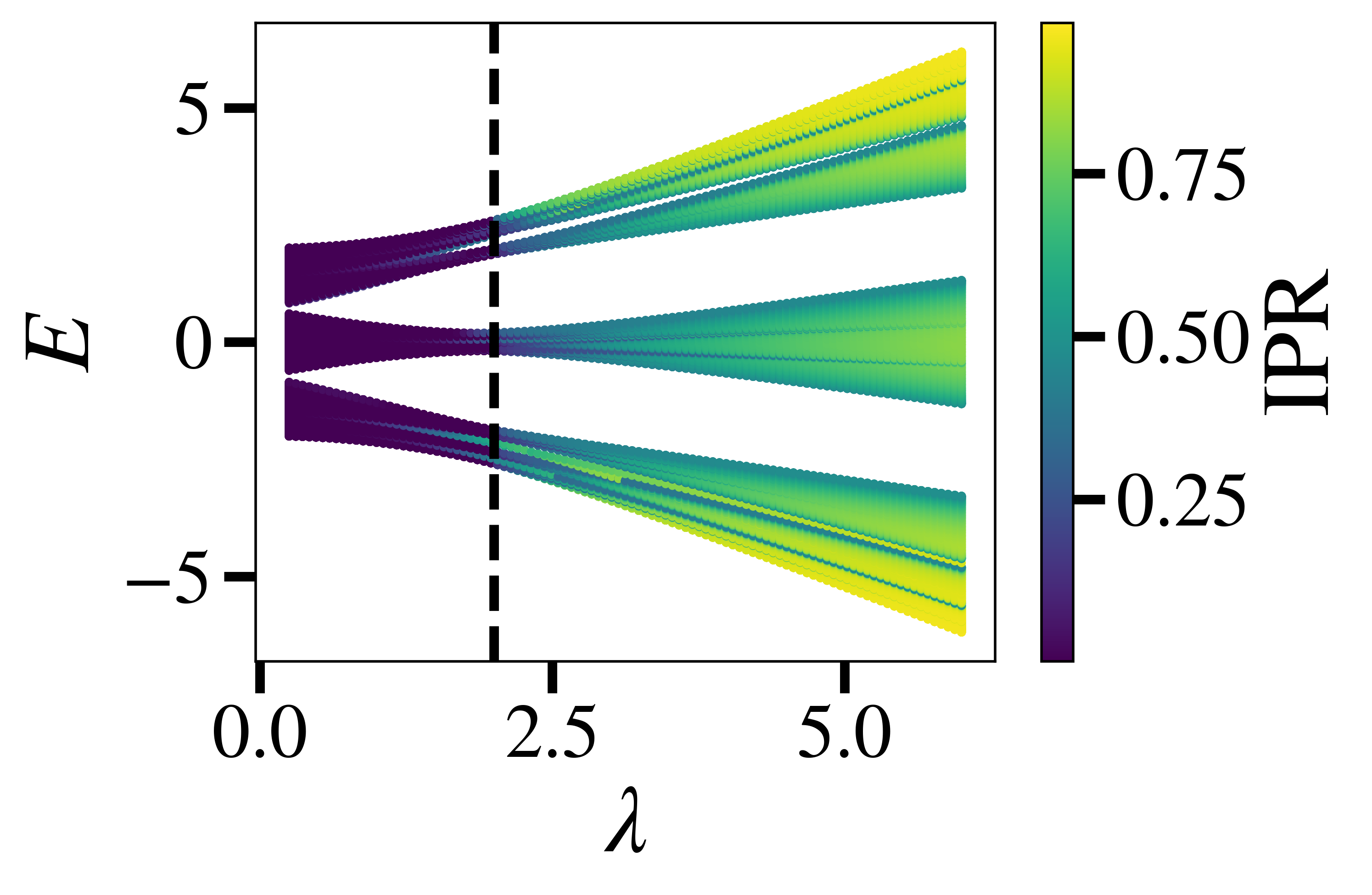}
        \caption{IPR as a function of $E$ and $\lambda$ for $q=1$.}
    \end{subfigure}
    \hspace{0.05\linewidth}
    \begin{subfigure}[t]{0.45\linewidth}
        \includegraphics[width=\linewidth]{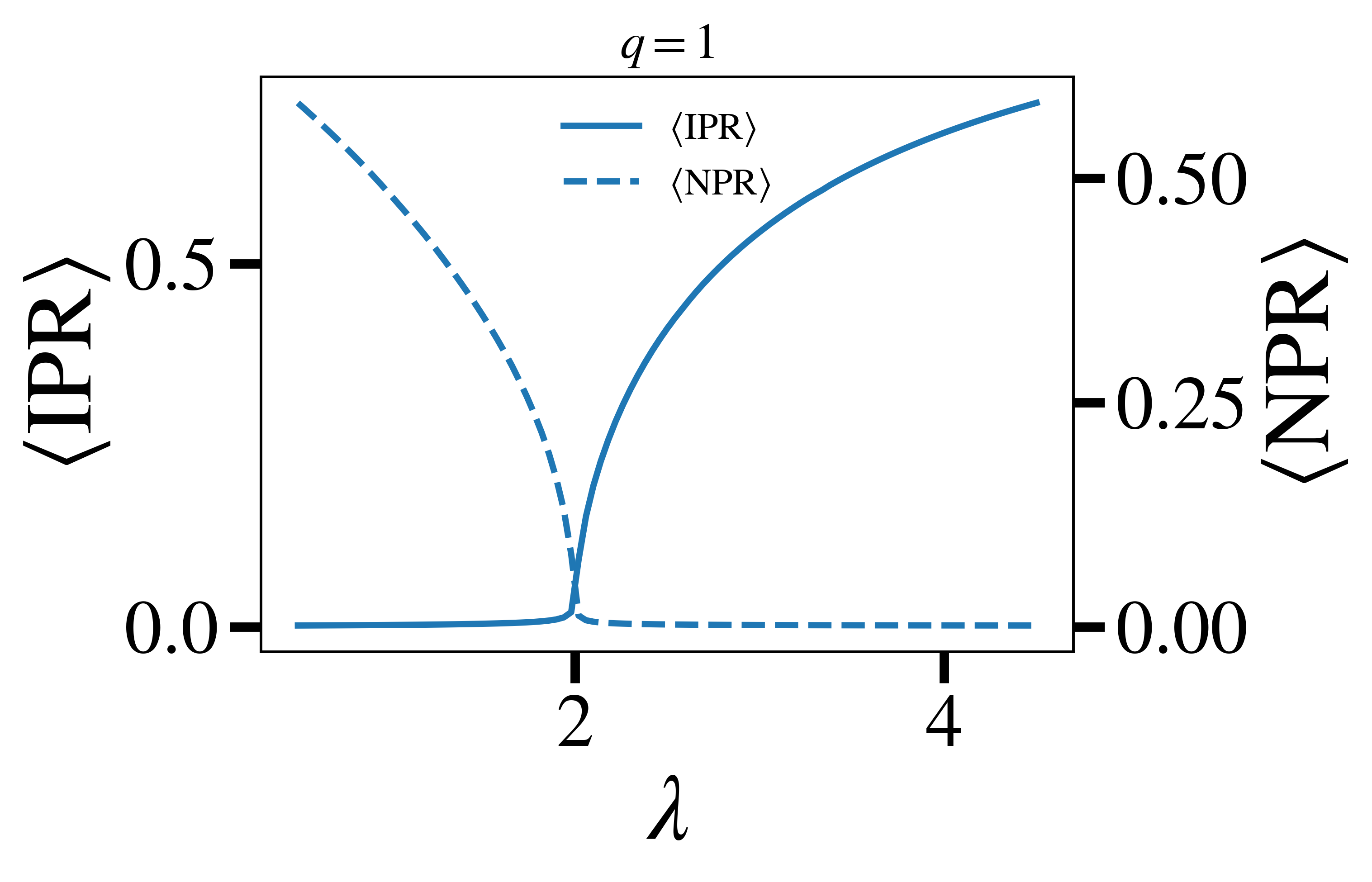}
        \caption{$\langle IPR \rangle$, $\langle NPR \rangle$ as a function of $\lambda$ at $q=1$.}
    \end{subfigure}
    \caption{(a)Energy resolved IPR and (b)$\langle IPR \rangle$, $\langle NPR \rangle$ as a function of $\lambda$ at $q=1$ for system size $N=1000$. }
    \label{fig:fig4}
\end{figure*}

\begin{figure*}
    \centering
    \includegraphics[width=1\textwidth]{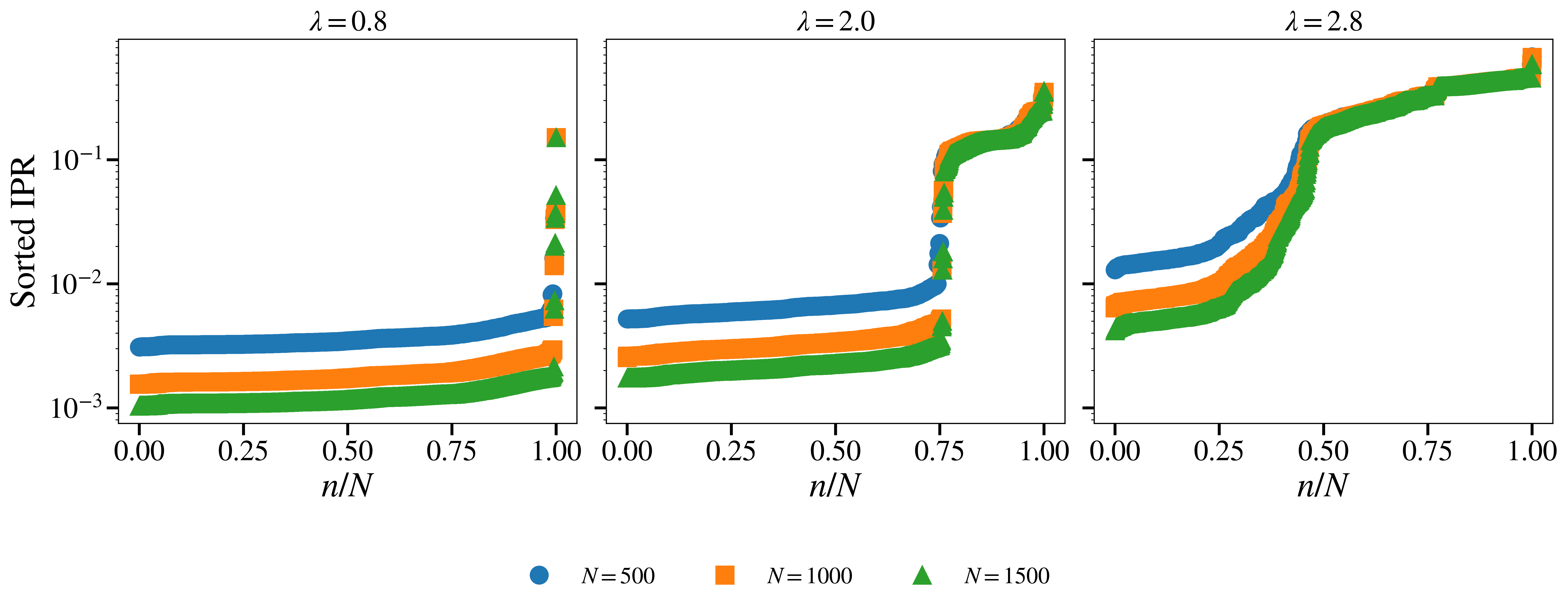}
    \caption{Sorted IPR as a function of normalized state index $\frac{n}{N}$ plotted for $N=500, ~1000$ and $1500$ for $\lambda=0.8,~ 2.0$ and $2.8$ respectively for $q=0.5$.}
    \label{fig:fig5}
\end{figure*}

\subsection{Emergent correlated hopping hierarchy}
\label{sec:hopping_hierarchy}

The kinetic deformation in Eq.~(\ref{eq:qkinetic}) has a useful real-space interpretation that gives rise to an explicit departure from the conventional Aubry-Andr\'e model. The deformed kinetic operator can be expanded as:
\begin{equation}
H_{\rm kin}^{(q)}= -t\frac{\sinh(\eta X)}{\sinh\eta}
= -t\sum_{m=0}^{\infty} \frac{\eta^{2m+1}} {(2m+1)!\sinh\eta}
X^{2m+1},
\label{eq:kinetic_expansion}
\end{equation}
where $q=e^\eta$. Since powers of $X=T+T^\dagger$ generate hopping over increasing lattice distances, Eq.~(\ref{eq:kinetic_expansion}) represents an infinite set of odd-range hopping processes. In particular, the kinetic Hamiltonian can be written as
\begin{equation}
H_{\rm kin}^{(q)} = -\sum_{\ell=0}^{\infty}
t_{2\ell+1}(q) \left( T^{2\ell+1}+T^{-(2\ell+1)} \right),
\label{eq:hierarchy}
\end{equation}
where the hopping amplitudes $t_{2\ell+1}(q)$ are completely determined by the single deformation parameter $q$. This structure is qualitatively different from an extended Aubry-Andr\'e model in which several longer-range hopping amplitudes are introduced as independent parameters. Here, the amplitudes of all hopping channels are algebraically correlated and are generated simultaneously by the same kinetic deformation. For example, expanding around the undeformed limit $\eta=0$ gives
\begin{align}
t_1(q)&=
t\left[
1+\frac{\eta^2}{3}
+\frac{\eta^4}{45}
+\mathcal{O}(\eta^6)
\right],
\label{eq:t1_expansion}\\
t_3(q)&=
t\left[
\frac{\eta^2}{6}
+\frac{\eta^4}{30}
+\mathcal{O}(\eta^6)
\right],
\label{eq:t3_expansion}\\
t_5(q)&=
t\left[
\frac{\eta^4}{120}
+\mathcal{O}(\eta^6)
\right],
\label{eq:t5_expansion}
\end{align}
and higher odd-range hopping amplitudes appear successively at higher orders in $\eta$. Thus, sufficiently close to $q=1$, the leading correction to nearest-neighbor hopping is a $3-$rd neighbor process, while $5-$th, $7-$th, and higher odd-range processes form a systematic all-orders hierarchy.

The absence of even-range hopping is a direct consequence of the odd functional dependence of the kinetic operator on $X$. More generally, the coefficient of a hopping process at separation $2\ell+1$ receives contributions from all higher odd powers of $X$ in Eq.~(\ref{eq:kinetic_expansion}). Consequently, the hierarchy is not a finite-range truncation but an exact representation of the deformed kinetic operator. The conventional Aubry-Andr\'e model is recovered smoothly in the limit $\eta\rightarrow0$ ($q\rightarrow1$), for which
\begin{equation}
t_1\rightarrow t,
\qquad
t_{2\ell+1}\rightarrow0
\quad(\ell\geq1).
\end{equation}
Hence, the deformation continuously connects the self-dual nearest-neighbor Aubry-Andr\'e model to a model containing an infinite set of correlated kinetic processes. The same hierarchy is directly visible in momentum space. For a translation-invariant kinetic operator, the exact single-particle dispersion is
\begin{equation}
\varepsilon_q(k)= -t\frac{\sinh(2\eta\cos k)} {\sinh\eta}.
\label{eq:exact_dispersion}
\end{equation}
Equivalently, its Fourier representation can be expressed as
\begin{equation}
\varepsilon_q(k)=
-2\sum_{\ell=0}^{\infty}
t_{2\ell+1}(q)
\cos\!\left[(2\ell+1)k\right].
\label{eq:dispersion_hierarchy}
\end{equation}
Thus, the nonlinear deformation of the dispersion and the infinite odd-range hopping hierarchy are two equivalent descriptions of the same kinetic modification. At $q=1$, the dispersion reduces to the single cosine $\varepsilon(k)=-2t\cos k$, whereas for $q\neq1$ higher odd Fourier harmonics are generated in a controlled manner. This observation is central to the localization problem considered below. The quasi-periodic potential $V_n=\lambda\cos(2\pi\alpha n+\phi),$ itself is left unchanged. Instead, the entire modification of the kinetic sector is encoded in the single parameter $q$. The resulting correlated hopping hierarchy therefore provides a controlled route for departing from the nearest-neighbor, self-dual Aubry-Andr\'e limit while preserving the same real-space quasi-periodic potential. As discussed below, this departure from the single-cosine kinetic structure is accompanied by the breaking of the standard Aubry-Andr\'e self-duality and the emergence of energy-dependent localization properties.

\subsection{Deformation-induced modification of the group velocity}
\label{sec:group_velocity}

The kinetic deformation also modifies the single-particle velocity through the nonlinear dispersion in Eq.~(\ref{eq:exact_dispersion}). For the translation-invariant kinetic sector, the group velocity is obtained from
\begin{equation}
v_q(k) = \frac{\partial \varepsilon_q(k)}{\partial k},
\end{equation}
which gives
\begin{equation}
v_q(k) = \frac{2t\eta}{\sinh\eta} \sin k\, \cosh(2\eta\cos k), \qquad
\eta=\ln q.
\label{eq:group_velocity}
\end{equation}
In the undeformed limit, $\eta\rightarrow0$, this expression continuously reduces to the standard nearest-neighbor result
\begin{equation}
v_{q=1}(k)=2t\sin k.
\end{equation}

For $q\neq1$, the factor $\cosh(2\eta\cos k)$ introduces a momentum-dependent modification that cannot be represented by a simple overall renormalization of the nearest-neighbor hopping. Consequently, the deformation changes not only the bandwidth but also the momentum dependence of the quasiparticle velocity. The result can be seen in Fig.~\ref{fig:kinetic_deformation}(a) and Fig.~\ref{fig:kinetic_deformation}(b). As the deformation is increased, the velocity profile develops additional structure away from the conventional maximum at $k=\pi/2$ (see Fig.~\ref{fig:kinetic_deformation}(b)). In particular, for sufficiently strong deformation, the momentum at which the maximum velocity occurs is shifted away from $k=\pi/2$, reflecting the higher harmonics generated by the kinetic deformation. The evolution of the velocity profile is directly connected to the emergent hopping hierarchy discussed above. Writing the dispersion as
\begin{equation}
\varepsilon_q(k) = -2\sum_{\ell=0}^{\infty} t_{2\ell+1}(q) \cos[(2\ell+1)k],
\end{equation}
the corresponding velocity can equivalently be expressed as
\begin{equation}
v_q(k) = 2\sum_{\ell=0}^{\infty} (2\ell+1)t_{2\ell+1}(q) \sin[(2\ell+1)k].
\label{eq:velocity_hierarchy}
\end{equation}
Thus, the deformation-induced modification of the velocity provides a direct momentum-space manifestation of the correlated odd-range hopping hierarchy in real space. The same single parameter $q$ controls both the relative strength of the higher-range hopping processes and the higher-harmonic structure of the group velocity. Thus, the kinetic deformation alone is responsible for the altered dispersion and velocity structure. This provides a complementary view of the kinetic deformation before its consequences for localization and the breaking of Aubry-Andr\'e self-duality are considered.
\subsection{The $q$-deformed Aubry-Andr\'e Hamiltonian}

Combining Eq.~(\ref{eq:qkinetic}) with the quasiperiodic potential, we define
\begin{equation}
 H_q = -t\frac{\sinh[\eta(T+T^\dagger)]}{\sinh\eta} + \lambda \cos(2\pi\alpha N+\phi).
\label{eq:fullq}
\end{equation}
Which for small $\eta$ takes the form in Eq.~\ref{eq:kinreal}
\begin{align}
H_q=&-t_1\sum_n(c_{n+1}^\dagger c_n+\mathrm{H.c.})\nonumber
\\&-t_3\sum_n (c_{n+3}^\dagger c_n+\mathrm{H.c.}) \nonumber
\\ & +\lambda\sum_n\cos(2\pi\alpha n+\phi)c_n^\dagger c_n +O(\eta^4),
\end{align}
where
\begin{equation}
t_1 = t\left(1+\frac{\eta^2}{3}\right), \qquad
t_3=\frac{t\eta^2}{6}.
\label{eq:t13}
\end{equation}
The resulting Hamiltonian may therefore be interpreted as a translationally invariant lattice with correlated odd-distance tunneling processes, whose relative amplitudes are controlled by a single parameter \(q\). The corresponding single-particle Schr\"odinger equation is
\begin{align}
E\psi_n =& -t_1(\psi_{n+1}+\psi_{n-1}) \nonumber\\
&-t_3(\psi_{n+3}+\psi_{n-3}) \nonumber
\\ & +\lambda \cos(2\pi\alpha n+\phi)\psi_n +O(\eta^4).
\label{eq:singleparticle}
\end{align}

It is immediately apparent that
\begin{equation}
    \lim_{q\rightarrow1}t_1=t,
    \qquad
    \lim_{q\rightarrow1}t_3=0.
\end{equation}
Consequently,
\begin{equation}
\lim_{q\rightarrow1}H_q=H_{\rm AA}.
\label{eq:AAlimit}
\end{equation}

\subsection{Symmetry under $q \to q^{-1}$}

An important property of the kinetic deformation introduced above is its invariance under the transformation $q\to q^{-1}$. Since $q=e^\eta$, this transformation is equivalent to
\begin{equation}
\eta\longrightarrow-\eta.
\end{equation}
The deformed kinetic operator is
\begin{equation}
H_{\mathrm{kin}}^{(q)}
=-t\frac{\sinh\left[\eta(T+T^\dagger)\right]}
{\sinh\eta}.
\end{equation}
Under $\eta\rightarrow-\eta$, both the numerator and denominator change sign,
\begin{align}
\sinh\left[-\eta(T+T^\dagger)\right]&=
-\sinh\left[\eta(T+T^\dagger)\right], \notag
\\ & \text{ and } \sinh(-\eta) = -\sinh\eta.
\end{align}
Consequently,
\begin{equation}
H_{\mathrm{kin}}^{(q^{-1})}= -t\frac{\sinh\left[-\eta(T+T^\dagger)\right]}{\sinh(-\eta)}= H_{\mathrm{kin}}^{(q)}.
\end{equation}
Because the quasi-periodic potential is independent of $q$, the complete Hamiltonian satisfies $H(q)=H(q^{-1})$. Thus, $q$ and $1/q$ correspond to exactly the same Hamiltonian and, consequently, have identical spectra and eigenstates. This symmetry implies that the deformation is controlled by $|\ln q|$, rather than by $q$ itself. The point $q=1$ is the undeformed Aubry-Andr\'e limit, while the parameter regions $q>1$ and $0<q<1$ are physically identical.

\subsection{Momentum-space representation}

Since
\begin{equation}
    X\ket{k}=2\cos k\ket{k},
\end{equation}
the exact $q$-deformed kinetic dispersion is
\begin{equation}
\epsilon_q(k)=-t\frac{\sinh(2\eta\cos k)}{\sinh\eta}.
\label{eq:qdispersion}
\end{equation}

For $\eta\rightarrow0$,
\begin{equation}
\epsilon_q(k) \rightarrow -2t\cos k.
\end{equation}

Expanding Eq.~(\ref{eq:qdispersion}),
\begin{align}
\epsilon_q(k)=&-2t\cos k-\frac{4t\eta^2}{3}\cos^3 k
+\frac{t\eta^2}{3}\cos k+O(\eta^4).
\end{align}

Using $\cos^3k=\frac{3\cos k+\cos3k}{4},$ we obtain

\begin{equation}
\epsilon_q(k)=-2t\left(1+\frac{\eta^2}{3}\right)\cos k
-\frac{t\eta^2}{3}\cos3k+O(\eta^4).  
\label{eq:qdispersionexp}
\end{equation}

This is consistent with the real-space hopping amplitudes in Eq.~(\ref{eq:t13}).

\subsection{Breaking of Aubry duality}

The localization transition of the conventional AA model is intimately related to its exact self-duality. To examine how the $q$-deformation modifies this property, we introduce the Aubry dual transformation
\begin{equation}
\psi_n=\sum_m e^{i2\pi\alpha nm}\phi_m.
\label{eq:dualitytransform}
\end{equation}

For a hopping term of range $r$,
\begin{equation}
\psi_{n+r}+\psi_{n-r} \to 2\cos(2\pi\alpha rm)\phi_m.
\end{equation}

Consequently, Eq.~(\ref{eq:singleparticle}) becomes
\begin{align}
E\phi_m =& -2t_1 \cos(2\pi\alpha m)\phi_m
\nonumber\\ & -2t_3\cos(6\pi\alpha m)\phi_m
\nonumber\\ &+ \frac{\lambda}{2} (\phi_{m+1}+\phi_{m-1})
+O(\eta^4).
\label{eq:dual}
\end{align}

At $q=1$,
\begin{equation}
t_1=t,\qquad t_3=0,
\end{equation}
and the usual AA dual Hamiltonian is recovered. For $q\neq1$, however, the dual quasi-periodic potential contains an additional harmonic, $\cos(6\pi\alpha m)$, which demonstrates that the exact self-duality of the AA model is broken by the $q$-deformation.

\begin{figure*}
    \centering
    \includegraphics[width=1\textwidth]{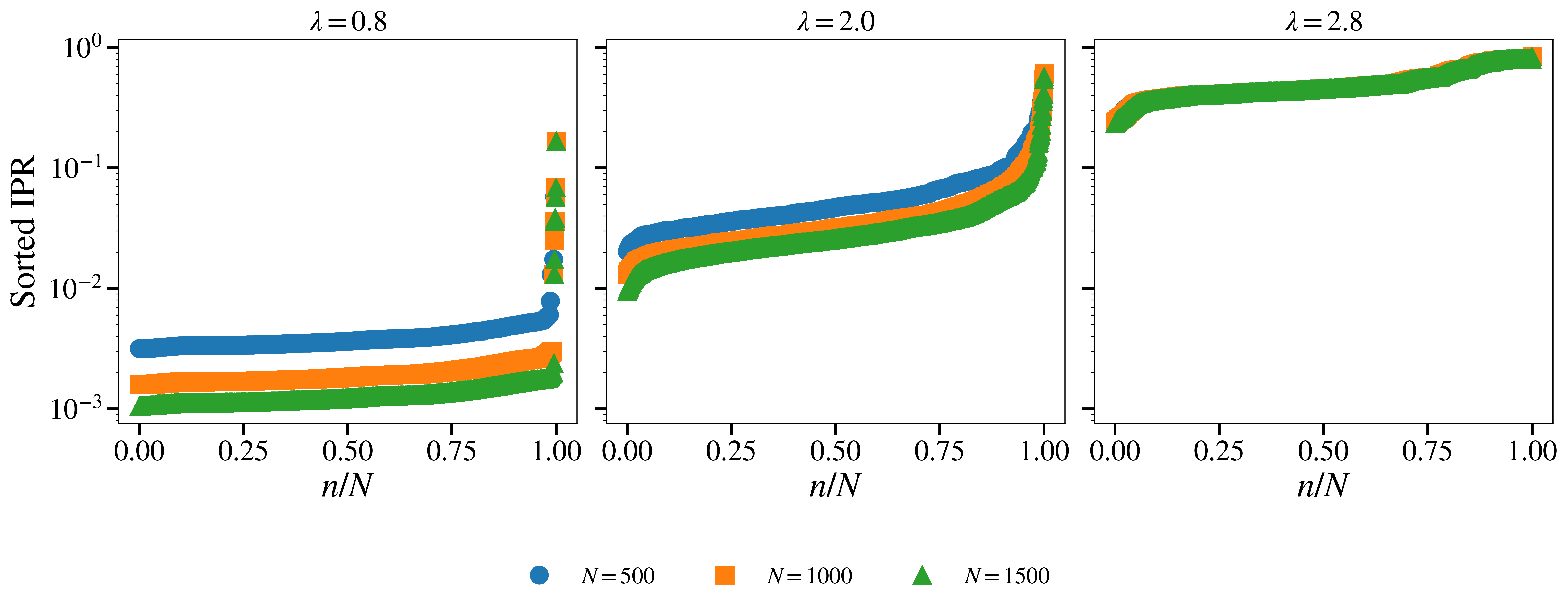}
    \caption{Sorted IPR as a function of normalized state index $\frac{n}{N}$ plotted for $N=500, ~1000$ and $1500$ for $\lambda=0.8,~ 2.0$ and $2.8$ respectively for $q=1$.}
    \label{fig:fig6}
\end{figure*}

\begin{figure*}
    \centering
    \includegraphics[width=1\textwidth]{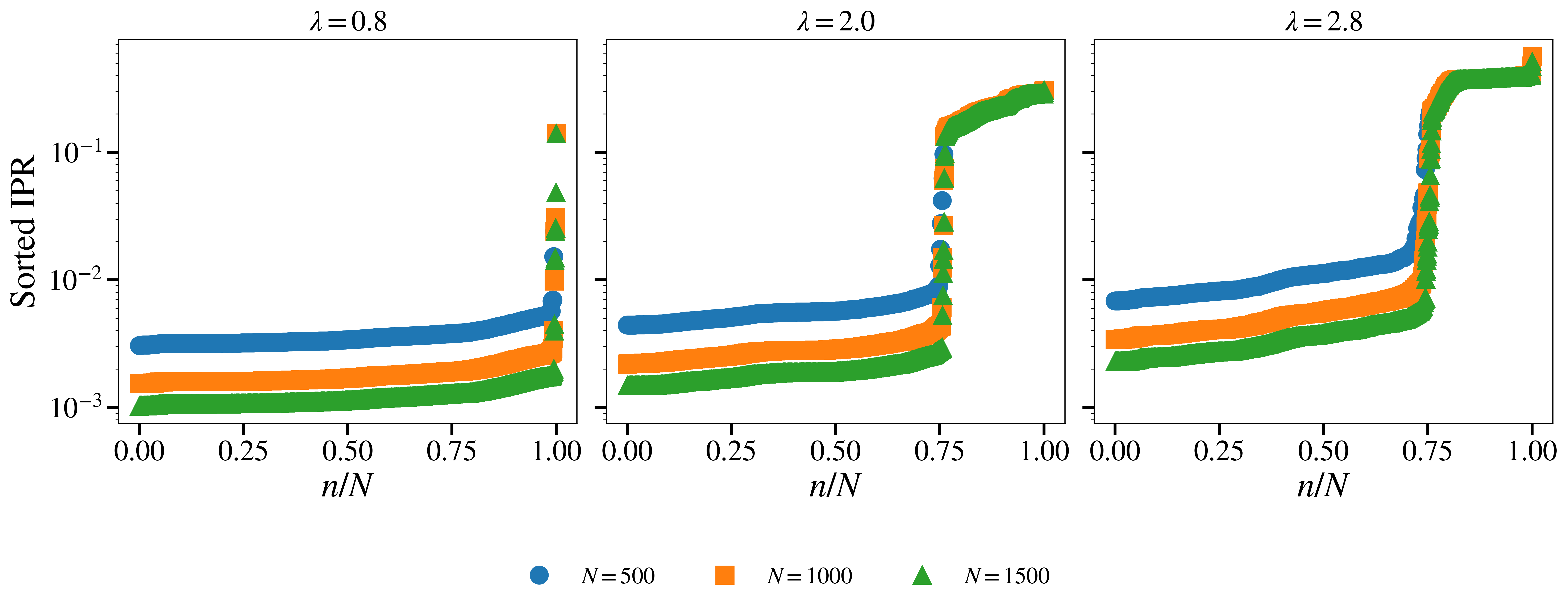}
    \caption{Sorted IPR as a function of normalized state index $\frac{n}{N}$ plotted for $N=500, ~1000$ and $1500$ for $\lambda=0.8,~ 2.0$ and $2.8$ respectively for $q=2.5$.}
    \label{fig:fig7}
\end{figure*}
\begin{figure}
    \centering
    \includegraphics[width=0.45\textwidth]{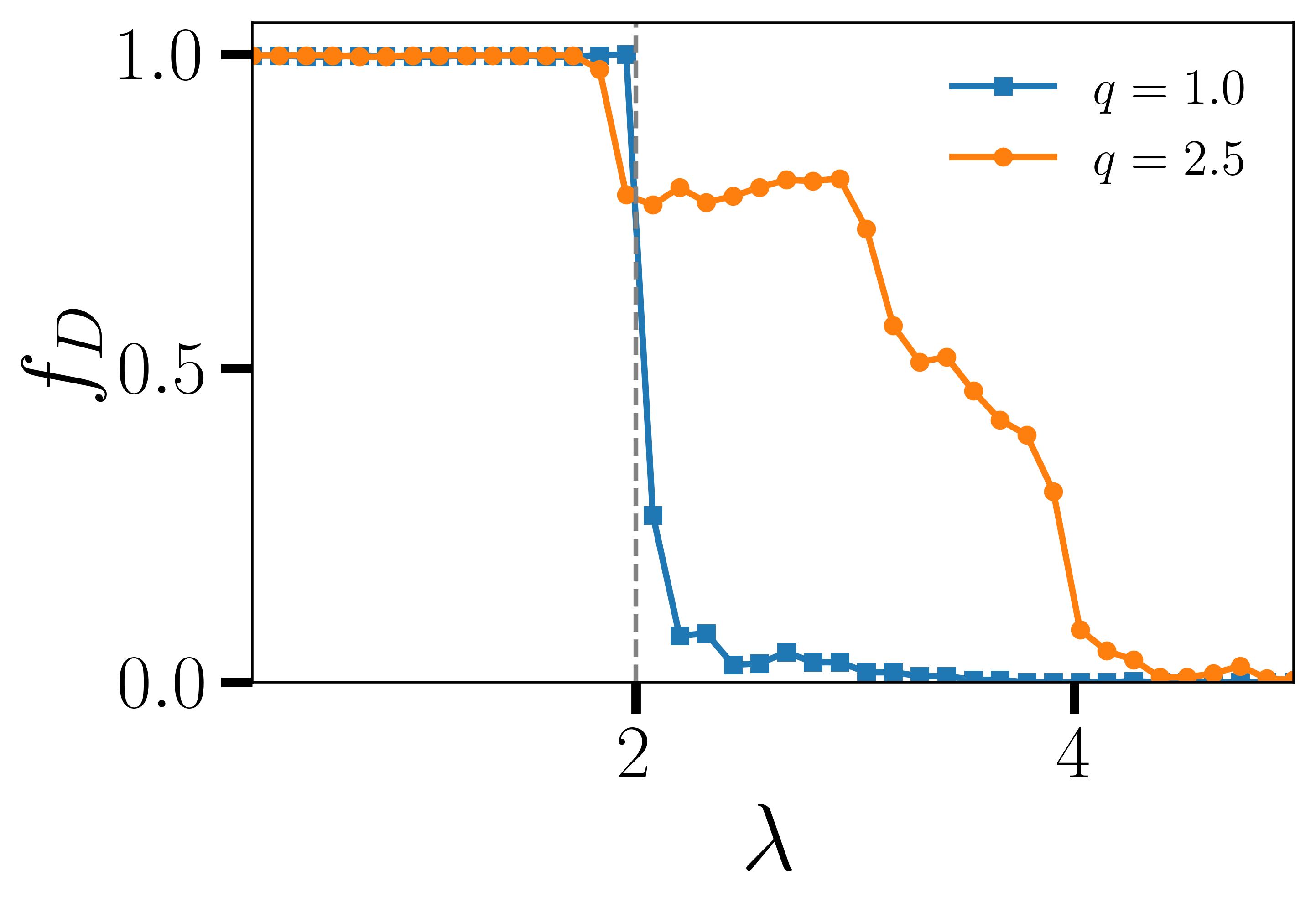}
    \caption{The plot of $f_D$ as a function of $\lambda$ for $q=1$ and $2.5$.}
    \label{frac_del}
\end{figure}

\section{Numerical results} \label{sec_numeric}
For our analysis, we took the exact form of the q-deformed Hamiltonian in Eq.~\ref{eq:fullq}. Now, to study the localization property, we first consider the inverse participation ratio (IPR), which measures how localized a state is. For the $k$th normalized eigenket, it is defined as \cite{ipr_1, ipr_2, ipr_3}:
\begin{equation} \label{ipr}
    IPR^{(k)} = \sum_{j=1}^{N} |\braket{j}{\xi_k}|^4.
\end{equation}
We recall that $0\le IPR^{(k)}\le 1$. A higher IPR value, $IPR^{(k)} = O(1)$, indicates that the corresponding eigenket is localized, while a lower value ($\sim 1/N$) suggests that the corresponding state is delocalized. Sometimes, the average IPR is calculated to assess the system's overall localization properties. For a system with $N$ normalized eigenkets which can be defined as:
\begin{equation} \label{ipr_av}
    \langle IPR \rangle = \frac{1}{N}\sum_{k=1}^{N} IPR^{(k)}. 
\end{equation}
A delocalized phase is characterized by $\langle IPR \rangle=0$ in the limit $N\to \infty$.

\par The participation ratio (PR) is a complementary, sometimes more convenient, measure of localization. We define the average PR in the following way:
\begin{equation}\label{pr_av}
    \langle PR \rangle = \frac{1}{N} \sum_{k=1}^{N} \frac{1}{IPR^{(k)}}.
\end{equation}

To gain more insight into the phases of a system, sometimes the normalized participation ratio (NPR) is also calculated. For the $k$th normalized eigenket, it is defined as 
\begin{equation}\label{npr}
    NPR^{(k)} = \frac{PR^{(k)}}{N},
\end{equation}
where $PR^{(k)}=\frac{1}{IPR^{(k)}}$ (see Eq. \ref{ipr}). The average of this quantity over all eigenkets provides a good understanding of the overall localization behavior of a system. The average NPR is calculated following,
\begin{equation}\label{npr_av}
    \langle NPR\rangle = \frac{1}{N}\sum_{k=1}^{N} NPR^{(k)} = \frac{\langle PR \rangle}{N},
\end{equation}
where $\langle PR \rangle$ is defined in Eq. \ref{pr_av}.
In the limit $N\to \infty$, $\langle NPR\rangle \sim \mathcal O(1)$ for the delocalized phase and  $\langle NPR\rangle \sim 0$ for the localized phase. 

Following Refs. \cite{Li2020, QP_4, goswami2026} and other related works, we call a phase delocalized if $\langle IPR \rangle = 0$ and $\langle NPR \rangle > 0$, and designate a phase localized if $\langle IPR \rangle > 0$ and $\langle NPR \rangle = 0$. Interestingly, there are some parameter regimes where both $\langle IPR \rangle > 0$ and $\langle NPR \rangle > 0$. This phase is called the mixed or intermediate phase. We note here that the crossover lines $c_1$ and $c_2$ in our phase diagram, Fig. \ref{phase_diag}, are determined and discussed in the following sections. The symmetric nature of the phase diagram about the $q=1$ point comes from the fact that the Hamiltonian $H_q$ is symmetric under $q \to q^{-1}$ transformation.

\subsection{Average IPR Phase Map}

Figure~\ref{fig:contour_ipr_q_lambda} shows the dependence of average IPR on both the potential strength and the kinetic deformation. For $q = 1$, the system reproduces the standard Aubry-Andr\'e limit, where the localization transition occurs sharply at $\lambda = 2$. As $q$ deviates from unity, the transition boundary broadens, indicating that the kinetic deformation modifies the localization behavior. The region to the left of the dashed line represents the delocalized phase, while the bright region at larger $\lambda$ and $q$ corresponds to the localized regime. The intermediate region between them highlights the emergence of mixed states with finite $\langle \mathrm{IPR} \rangle$ and $\langle \mathrm{NPR} \rangle$, revealing the deformation-induced departure from the conventional AA transition and has been shown explicitly in Fig.~\ref{phase_diag}. As a consistency check, we have verified that the average IPR and NPR curves converge well as the system size increases. We present the plots corresponding to $q=2.5$, as shown in Appendix~\ref{fig:appendix_convergence}. We find the phase boundaries $c_1$ and $c_2$ of the localization map in Fig.~\ref{phase_diag} with the help of these $\langle \mathrm{IPR} \rangle$ and $\langle \mathrm{NPR} \rangle$ plots, where $c_1$ separates delocalized (D) and mixed phase (M), and $c_2$ separates mixed (M) and localized phase (L). We explicitly calculated the points for $c_1$ and $c_2$ lines for different $q$ values. The points are extracted in the following way: for a given value of $q$, the $\lambda$ point before which the value of $\langle IPR \rangle \sim \mathcal O(1/L)$ and beyond which $\langle IPR \rangle$ is finite is considered for the $c_1$ line. Similarly, the $\lambda$ point beyond which the value of $\langle NPR \rangle \sim \mathcal O(1/L)$ and before which $\langle IPR \rangle$ is finite is considered for the $c_2$ line. The stability of the mixed phase with increasing system size is confirmed by the convergence plot in Appendix~\ref{fig:appendix_convergence}.

\subsection{$q$-dependence of the localization transition}

We next investigate the localization properties as a function of the deformation parameter $q$. First, we plot the energy resolved $\mathrm{IPR}(E,\lambda,q)$ as a function of $\lambda$ for $q=0.5$ and $q=2.5$ respectively, the corresponding results can be seen in Fig.~\ref{fig:fig2}(a) and (b) respectively. The calculation has been performed over the entire spectrum to see whether the critical quasiperiodic potential strength depends on energy. The plots show that for both $q$-values, the upper and lower bands remain delocalized well beyond $\lambda=2$, unlike in the case of $q=1$. Further, to understand the criticality of the states, we plot the average IPR and NPR over all the states. For our analysis, we choose two $q$ values: $q=0.5,$ $2.5$, to understand the deviation from the normal Aubrey Andr\'e limit at $q \to 1$. From Fig.~\ref{fig:fig3}(a), we see that for $q=0.5$ within the limit $2 < \lambda <3$ both $\langle IPR \rangle$  and $\langle NPR \rangle$ are finite, signals the mixed phase region. However, for $\lambda <2$, $\langle IPR \rangle=0$ and for $\lambda > 3 $, $\langle NPR \rangle=0$, representing a fully delocalized (low-IPR) and a fully localized (high-IPR) state, respectively. In a similar spirit, we can see in Fig.~\ref{fig:fig3}(b) at $q = 2.5$, between $1.8< \lambda < 4$, both $\langle IPR \rangle$  and $\langle NPR \rangle$ are finite, representing an intermediate phase. Also, we see for $\lambda< 1.8$, $\langle IPR \rangle =0$ and for $\lambda>4 $, $\langle NPR \rangle=0$ representing delocalized and localized phases respectively. This mixed phase is absent in the standard AA model and arises solely due to kinetic deformation. The presence of this regime highlights the role of kinetic deformation in generating unconventional localization phenomena without any direct or effective modification of the quasiperiodic potential.

\subsection{Recovery of the AA transition}

Now we also verify that the model reproduces the known AA localization transition at $q=1$. The transition is expected at $\lambda=2.$ From Fig.~\ref{fig:fig4}(a), we see a sharp transition at $\lambda=2$ with no extended states beyond that. Similarly, the $\langle IPR \rangle$ and $\langle NPR \rangle$ in Fig.~\ref{fig:fig4}(b) give complementary confirmation as there is no region with finite $\langle NPR \rangle$ and $\langle IPR \rangle$ that corresponds to the intermediate region.
\subsection{Sorted IPR Distributions and Fraction of delocalized States}

To gain a more detailed insight into the spectral organization of localized and extended states, we analyze the \textit{sorted inverse participation ratio (IPR)} as a function of the normalized state index $\frac{n}{N}$. Figures~\ref{fig:fig5}-\ref{fig:fig7} present the sorted IPR curves for different system sizes ($N = 500, 1000, 1500$) and quasi-periodic potential strengths at representative values of the deformation parameter $q = 0.5-2.5$.  

For the conventional AA limit ($q = 1$), the sorted IPR curves exhibit a sharp crossover: below the critical potential strength $\lambda_c = 2$, the majority of states display vanishing IPR and scales as $\mathcal O \left(1/N\right)$ consistent with delocalization, while above $\lambda_c$ nearly all states acquire finite IPR values of order $\mathcal O(1)$, and collapses for different system sizes indicating localization. This behavior confirms the energy-independent nature of the AA transition.  

In contrast, for $q = 0.5$ and $q = 2.5$, the sorted IPR curves reveal an interesting feature. For very small quasi-periodic potential strength $\lambda=0.8$, all the states have negligibly small IPR and follow scaling of the form: $\mathcal O\left(1/N\right)$. At intermediate quasi-periodic strengths, in the entire spectra, a finite fraction of states remain delocalized (IPR $\mathcal O \sim 1/N$) while others have a high size-independent value of IPR, with collapsed data for different $L$ representing localized (IPR $\sim \mathcal O(1)$) states, even within the same parameter regime. This coexistence directly signals the presence of intermediate (or mixed) states and the breakdown of universal criticality.  

We quantify this coexistence by computing the fraction of delocalized states, defined from the above sorted IPR plots of the entire spectra. We see for $q=0.5$ (see Fig.~\ref{fig:fig5}) and $q=2.5$ (see Fig.~\ref{fig:fig7}), the sorted spectra show a special nature; there is a clear distinction between the localized and delocalized states in that for $q=0.5$ (see Fig.~\ref{fig:fig5}) and $q=2.5$ (see Fig.~\ref{fig:fig7}), the sorted spectra exhibit a distinct character: there is a clear distinction between the localized and delocalized states across the entire spectra. A finite portion of the spectrum collapses across different system sizes, showing a size-independent IPR value indicating localized states, while the rest of the spectrum displays a systematic decrease in IPR with increasing system size. The fractional value along the $\frac{n}{N}$-axis for which the sorted IPR collapses for different system sizes is the fraction of delocalized states. We see that for both the $q$ values at $\lambda =0.8$, all the states are delocalized, but as we increase $\lambda$ ($2$ or $2.8$), the fraction decreases. However, we still get a finite fraction of delocalized states. In contrast to this observation, for $q=1$ (see Fig.~\ref{fig:fig6}) we see that all the states are localized as the system is at $\lambda=2.8$. Thus, these results provide direct numerical evidence of the presence of a fraction of delocalized states beyond the self-dual point of the standard AA model.

The fraction of delocalized states has been determined using the following method: For two different system sizes $N_1$ and $N_2$, we compute the sorted IPRs and then take the difference between the sorted IPR values for states with the same normalized state index. To measure this difference, we define the quantity $\mathcal D$, which can be defined as follows:
\begin{equation}
\mathcal{D} = \left|1-\frac{
\mathrm{IPR}^{\left(\frac{n_1}{N_1}\right)}_{N_1}}{
\mathrm{IPR}^{\left(\frac{n_2}{N_2}\right)}_{N_2}}\right|,
\quad \text{where} \quad \left(\frac{n_2}{N_2}\right)=\left(\frac{n_1}{N_1}\right).
\end{equation}
For $N_2=2N_1$, we have $n_2=2n_1$. For simplicity, we therefore choose $N_1=500$ and $N_2=1000$ in our numerical calculations. So, for a delocalized state $\mathrm{IPR}_{N_1} \sim \mathcal O(1/N_1)$ and $\mathrm{IPR}_{N_2} \sim \mathcal O(1/2N_1)$ and $\mathcal{D}=1$. On the other hand, when the system is fully localized, then $\mathrm{IPR}_{N_1} \sim \mathrm{IPR}_{N_2}\sim \mathcal O(1)$, as a consequence $\mathcal D=0$. To calculate the fraction of delocalized states, we consider the critical value of the normalized state index, $n_c/N$, at which the finite-size IPR measure $\mathcal{D}$ crosses the small threshold value $0.05$. The states satisfying $\mathcal{D}>0.05$ are identified as delocalized according to this criterion, and the corresponding fraction of delocalized states is defined as:
\begin{equation}
f_D=\frac{1}{N}\sum_{n=1}^{N}
\Theta\left(\mathcal{D}_n-0.05\right),
\end{equation}
where $\Theta(x)$ is the Heaviside step function. If $\mathcal{D}_n>0.05$ up to a critical state index $n_c$, this definition reduces to
\begin{equation}
f_D=\frac{n_c}{N}.
\end{equation}
We present the variation of $f_D$ as a function of $\lambda$ for $q=2.5$ and $q=1$. As shown in Fig.~\ref{frac_del}, for $q=1$, $f_D$ decreases sharply and vanishes at $\lambda=2$. In contrast, for $q=2.5$, $f_D$ remains finite beyond $\lambda=2$, indicating the presence of a mixed regime with localized and delocalized states. For both values of $q$, the fraction $f_D$ decreases with increasing $\lambda$.
\begin{figure}
\includegraphics[width=0.45\textwidth]{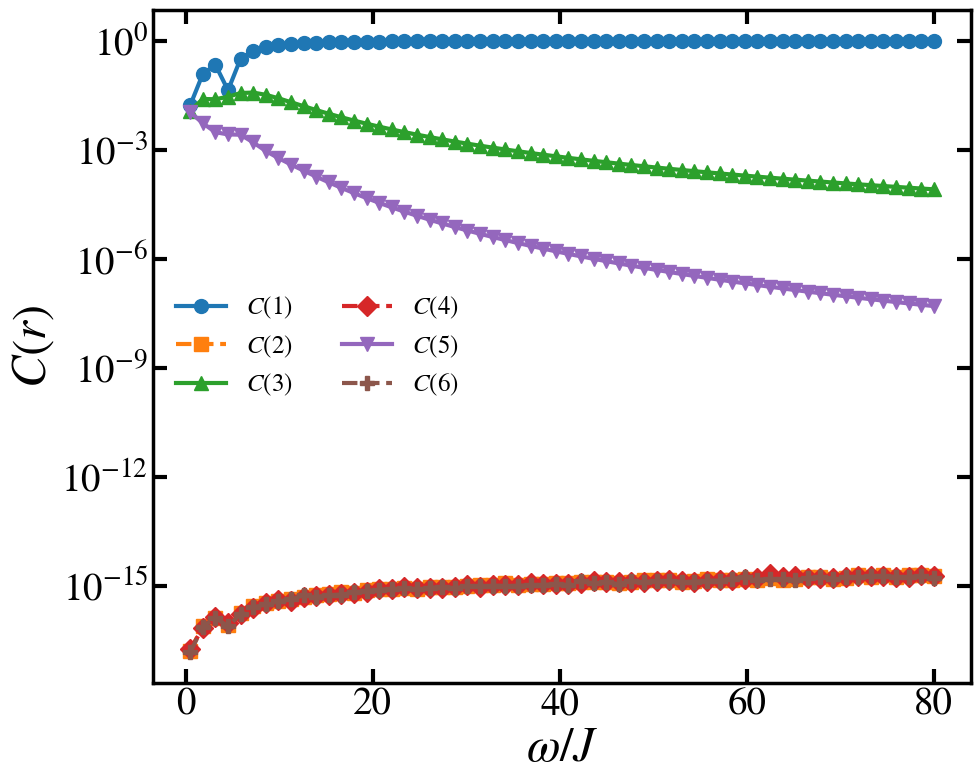}
\caption{Plot of odd and even hopping amplitudes as a function of $\omega$.}
\label{hopping_ampl}
\end{figure}
\section{Floquet interpretation of the kinetic deformation} \label{floqt_realization}
In this section, we establish a natural connection between the correlated long-range hopping hierarchy generated by the $q$-deformation and Floquet Hamiltonian engineering. To make this connection explicit, consider a tight-binding chain subjected to a periodic staggered square pulse,

\begin{align}
H(t) &=-J\sum_n(c_{n+1}^{\dagger}c_n+\mathrm{H.c.})
\\ &+A f(t)\sum_n(-1)^n n_n ,
\end{align}

where $f(t+T)=f(t)$ and $T=2\pi/\omega$. A transformation to the rotating frame removes the explicit driving potential and converts it into a time-dependent phase of the nearest-neighbor hopping,
\begin{equation}
H(t)=-J\left[e^{i\phi(t)P}T+
T^\dagger e^{-i\phi(t)P}\right],
\label{eq:floquet_clean_H}
\end{equation}
where
\begin{equation}
T=\sum_n |n+1\rangle\langle n|,
\qquad
P=\sum_n(-1)^n|n\rangle\langle n|.
\end{equation}
The operator $P$ satisfies the following relations:
\begin{equation}
P^2=1,\qquad
PT=-TP,\qquad
PT^\dagger=-T^\dagger P,
\end{equation}
while for a translationally invariant chain away from the boundary corrections,
\begin{equation}
TT^\dagger=T^\dagger T=1.
\end{equation}

We consider a zero-mean square force,
\begin{equation}
f(t)=\dot{\phi}(t)=
\begin{cases}
+A, & 0<t<T/2,\\
-A, & T/2<t<T,
\end{cases}
\qquad
\omega=\frac{2\pi}{T},
\label{eq:square_force}
\end{equation}
and choose $\phi(0)=0$. The resulting phase is triangular,
\begin{equation}
\phi(t)= \begin{cases}
At, & 0<t<T/2,\\
A(T-t), & T/2<t<T.
\end{cases}
\label{eq:triangular_phase}
\end{equation}
Introducing
\begin{equation}
x=\frac{\pi A}{\omega},
\end{equation}
the maximum value of the phase is $\phi_{\rm max}=x$. Since $P^2=1$, one may write
\begin{equation}
e^{i\phi(t)P}
=e^{i\phi(t)}\Pi_e+
e^{-i\phi(t)}\Pi_o,
\qquad
\Pi_{e,o}=\frac{1\pm P}{2}.
\end{equation}
We therefore expand
\begin{equation}
e^{i\phi(t)P}
=\sum_{m=-\infty}^{\infty}
U_m e^{im\omega t},
\qquad
U_m=u_m\Pi_e+v_m\Pi_o .
\label{eq:Um_definition}
\end{equation}
The Fourier coefficients are
\begin{align}
u_m &=\frac{1}{T}\int_0^Tdt\,
e^{i\phi(t)}e^{-im\omega t},
\\ &
v_m=\frac{1}{T}\int_0^Tdt\,
e^{-i\phi(t)}e^{-im\omega t}.
\end{align}
For the triangular phase in Eq.~\eqref{eq:triangular_phase}, these integrals give
\begin{equation}
u_m= \frac{ix\left[1-(-1)^m e^{ix}\right]}{x^2-\pi^2m^2},
\label{eq:um_exact}
\end{equation}
and
\begin{equation}
v_m= -\frac{ix\left[1-(-1)^m e^{-ix}\right]} {x^2-\pi^2m^2}.
\label{eq:vm_exact}
\end{equation}
For $m=0$,
\begin{equation}
u_0=e^{ix/2}\frac{2\sin(x/2)}{x},
\qquad
v_0=e^{-ix/2}\frac{2\sin(x/2)}{x}.
\label{eq:u0v0}
\end{equation}
Where the constant phase can be absorbed by a constant gauge transformation. These coefficients satisfy the following relations:
\begin{equation}
u_{-m}=u_m, \qquad
v_{-m}=v_m,\qquad
u_m^*=v_m,
\end{equation}
as required by the Hermiticity of the time-dependent Hamiltonian. The Fourier components of Eq.~\eqref{eq:floquet_clean_H} are
\begin{equation}
H_m=-J\left(U_mT+T^\dagger U_m^\dagger \right).
\label{eq:Hm}
\end{equation}
Using $PT=-TP$, the exact commutator between two Fourier components is
\begin{equation}
[H_m,H_n]=J^2 \left(u_mv_n-u_nv_m\right) P\left(T^2+T^{\dagger 2}+2\right).
\label{eq:first_commutator}
\end{equation}
Thus, at the level of operator structure, the first commutator generates next-nearest-neighbor hopping. However, for the symmetric square-force protocol considered here,
\begin{equation}
u_{-m}=u_m,\qquad v_{-m}=v_m,
\end{equation}
and consequently
\begin{equation}
[H_{-m},H_m]=0.
\label{eq:first_commutator_zero}
\end{equation}
The first-order van Vleck correction therefore vanishes,
\begin{equation}
H_F^{(1)}=\frac{1}{\omega} \sum_{m\neq0}
\frac{[H_{-m},H_m]}{2m}=0.
\label{eq:first_vV_zero}
\end{equation}

The leading nontrivial correction consequently occurs at order $\omega^{-2}$. We use the following van Vleck expansion expression:
\begin{align}
&H_F={}H_0+\frac{1}{\omega}
\sum_{m\neq0}\frac{[H_{-m},H_m]}{2m}
\nonumber\\&+\frac{1}{\omega^2}
\left[\sum_{m\neq0}
\frac{[H_{-m},[H_0,H_m]]}{2m^2}
+\sum_{\substack{m,n\neq0\\m+n\neq0}}
\frac{[H_{-m},[H_{m-n},H_n]]}{3mn}
\right] \notag \\ & +O(\omega^{-3}).
\label{eq:van_vleck}
\end{align}
Using Eq.~\eqref{eq:first_commutator}, the first nested commutator appearing at second order is
\begin{align}
[H_{-m},[H_0,H_m]]
={}&J^3 \left(u_0v_m-u_mv_0\right)
\nonumber\\&\times\left[
(u_m+v_m)P+(u_m-v_m)\right]
\nonumber\\&\times\left(
T^3+3T+3T^\dagger+T^{\dagger3}\right).
\label{eq:first_nested}
\end{align}
Similarly, the second nested commutator is
\begin{align}
[H_{-m},[H_{m-n},H_n]]
={}&J^3\left(u_{m-n}v_n-u_nv_{m-n}\right) \nonumber\\
&\times\left[(u_m+v_m)P+(u_m-v_m)
\right]\nonumber\\&\times
\left(T^3+3T+3T^\dagger+T^{\dagger3}
\right).
\label{eq:second_nested}
\end{align}
The appearance of the third-neighbor operator follows directly from
\begin{equation}
(T+T^\dagger)^3=T^3+3T+3T^\dagger+T^{\dagger3}.
\label{eq:S3}
\end{equation}
Consequently, the second-order Floquet Hamiltonian contains both a nearest-neighbor renormalization and a third-neighbor hopping contribution. For the first nested commutator, the substitution of the exact Fourier coefficients gives
\begin{equation}
u_0v_m-u_mv_0 =\frac{2i\left[(-1)^m-1\right]\sin x}
{x^2-\pi^2m^2},
\label{eq:umvm_combination}
\end{equation}
so that only odd Fourier harmonics contribute to this term. Furthermore,
\begin{equation}
u_m+v_m=\frac{2(-1)^m x\sin x}
{x^2-\pi^2m^2},
\label{eq:umplusvm}
\end{equation}
and
\begin{equation}
u_m-v_m =\frac{2ix\left[1-(-1)^m\cos x\right]}{x^2-\pi^2m^2}.
\label{eq:umminusvm}
\end{equation}
For odd $m$, Eq.~\eqref{eq:first_nested} therefore reduces to
\begin{align}
[H_{-m},[H_0,H_m]]={}&\frac{8J^3x\sin x}{\left(x^2-\pi^2m^2\right)^2}
\nonumber\\
&\times \left[i\sin x \, P-(1+\cos x)
\right]
\nonumber\\
&\times\left(
T^3+3T+3T^\dagger+T^{\dagger3}\right).
\label{eq:first_nested_simplified}
\end{align}

The double-sum contribution is similarly controlled by the exact combination
\begin{equation}
u_{m-n}v_n-u_nv_{m-n}=\frac{2i\sin x
\left[(-1)^n-(-1)^{m-n}\right]}
{\left[x^2-\pi^2(m-n)^2\right]
\left[x^2-\pi^2n^2\right]}.
\label{eq:double_sum_combination}
\end{equation}
In particular, this coefficient vanishes identically for even $m$. For odd $m$ it becomes
\begin{equation}
u_{m-n}v_n-u_nv_{m-n}
=\frac{4i(-1)^n\sin x}
{\left[x^2-\pi^2(m-n)^2\right]
\left[x^2-\pi^2n^2\right]}.
\end{equation}
So, Eq.~\ref{eq:second_nested}, reduces to:
\begin{align}
[H_{-m},[H_{m-n},H_n]]
={}&\frac{4ix(-1)^n\sin x}
{\left[x^2-\pi^2(m-n)^2\right]
\left[x^2-\pi^2n^2\right]} \nonumber\\
&\times \left[i\sin x \, P-(1+\cos x)
\right]\nonumber\\&\times
\left(T^3+3T+3T^\dagger+T^{\dagger3}
\right).
\label{eq:double_sum_odd}
\end{align}
\par 
The $i\sin x\, P-(1+\cos x)$ factor appearing in the second-order Floquet contribution can be written in exponential form as
\begin{equation}
i\sin x\,P-(1+\cos x) =
-2\cos\frac{x}{2}\,
e^{-iPx/2}.
\label{eq:phase_identity}
\end{equation}
where $e^{-iPx/2} =\cos\frac{x}{2}\,I -i\sin\frac{x}{2}\,P,$

Eq.~\eqref{eq:phase_identity} follows directly from $P^2=I$ and the half-angle identities. The factor $e^{-iPx/2}$ represents a site-dependent phase because
\begin{equation}
P\ket{n}=(-1)^n\ket{n},
\qquad
e^{-iPx/2}\ket{n} =
e^{-i(-1)^n x/2}\ket{n}.
\end{equation}
For the odd-range hopping operators generated here, $PT^r=-T^rP$ for odd $r$. Consequently, the local gauge transformation
\begin{equation}
G=e^{iPx/4}
\end{equation}
gives
\begin{equation}
G\left(e^{-iPx/2}T^r\right)G^\dagger
=T^r, \qquad r=1,3,5,\ldots .
\label{eq:gauge_remove}
\end{equation}
Thus, the site-dependent phase in Eq.~\eqref{eq:phase_identity} is a gauge-dependent hopping phase, and can be removed simultaneously from all odd-range hopping terms. The remaining scalar factor $2\cos(x/2)$ can be absorbed into the corresponding scalar hopping coefficient. Since this transformation is local and unitary, it does not change the quasi-energy spectrum or the localization properties. Eq.~\eqref{eq:first_nested_simplified}-\eqref{eq:double_sum_odd} demonstrates explicitly how a purely kinetic tight-binding Hamiltonian can generate longer-range hopping operators through the Floquet commutator hierarchy. No onsite potential is introduced in this construction. Collecting the surviving contributions up to second order in the high-frequency expansion, the effective Floquet Hamiltonian can be written in the compact form by taking

\[H_F^{(2)}=-\frac{J^3}{\omega^2} \alpha^{(2)}e^{-iPx/2} \left(T^3+3T+3T^\dagger+T^{\dagger3}
\right)\] where 
\begin{align}
\alpha^{(2)} &= \bigg( \sum_{m \ne 0} \frac{8x\sin x}{2m^2\left(x^2-\pi^2m^2\right)^2}
\nonumber+ \\ & \sum_{\substack{m,n\neq0\\m+n\neq0}} \frac{4ix(-1)^n\sin x}
{3mn\left[x^2-\pi^2(m-n)^2\right]\,
\left[x^2-\pi^2n^2\right]} \bigg) \notag \\&  \times 2\cos(x/2) \notag, \quad \text{and} \qquad \\& \alpha^{(0)} =\frac{2\sin (x/2)}{x}.
\end{align}
Thus, the resultant high-frequency expansion will look as follows:
\begin{align}
H_F = &-J\alpha^{(0)}\left(T+T^\dagger\right) \\ & -\frac{J^3}{\omega^2}\alpha^{(2)}e^{-iPx/2}\bigg( T^3+T^{\dagger 3} +3(T+T^\dagger)\bigg) 
\end{align}
where
\begin{equation}
T=\sum_n |n+1\rangle\langle n|,\qquad
T^3=\sum_n |n+3\rangle\langle n|.
\end{equation}


Thus, the Floquet drive generates an explicit third-neighbor hopping
term at order $\omega^{-2}$, providing a direct route to a longer-range kinetic hopping hierarchy. For analytical complexity we restrict our analysis to $\mathcal O(\omega^{-2})$
\par To confirm our claim, we do an exact numerical Floquet Hamiltonian calculation, and we plot the average amplitude of even and odd hopping amplitudes in Fig.~\ref{hopping_ampl}.
For a time-periodic Hamiltonian with period $T$, the stroboscopic evolution over one driving cycle is described by the Floquet operator~\cite{Floqute_theory_1, Floqute_theory_2, Floquet_singleparticle1, Floquet_singleparticle2}. To verify the emergence of longer-range hopping, we compute the exact Floquet operator over one driving period and obtain the effective Floquet Hamiltonian from the principal branch of the matrix logarithm,
\begin{equation}
H_F=\frac{i}{T}\log U_F(T).
\end{equation}
We then characterize the range of the effective Hamiltonian through the distance-resolved amplitudes
\begin{equation}
C(r)=\frac{1}{N-r}\sum_{n=1}^{N-r}
\left|(H_F)_{n,n+r}\right|.
\end{equation}
For $A=4$, $J=1$, and $N=500$, the exact Floquet Hamiltonian exhibits predominantly odd-range hopping over the entire-frequency regime. In particular, the $r=3$ and $r=5$ amplitudes are finite at finite frequency and decrease as $\omega$ increases, while the even-range amplitudes are strongly suppressed. This behavior is consistent with the high-frequency expansion, in which the first nontrivial correction occurs at order $\omega^{-2}$ and generates the third-neighbor hopping term explicitly. The odd-range amplitudes vanish progressively in the $\omega\rightarrow\infty$ limit, recovering the static nearest-neighbor tight-binding structure.

\section{Conclusion} \label{conc}
We have introduced a $q$-deformed generalization of the one-dimensional Aubry-Andr\'e model in which the deformation is applied directly to the lattice kinetic operator, while the quasiperiodic onsite potential is kept unchanged. The construction preserves Hermiticity and the uniform lattice structure and reduces exactly to the conventional Aubry-Andr\'e Hamiltonian in the limit $q\to1$. More broadly, the present framework provides a route to study localization beyond the standard Aubry-Andr\'e universality class via a controlled deformation of the kinetic energy. It also establishes a useful distinction from generic long-range quasiperiodic models: here, the complete hierarchy of long-range processes is generated by a single underlying deformation rather than by independently tuning multiple hopping amplitudes. The small $\eta$ expansion, with $\eta=\ln q$, makes this structure transparent: the leading correction generates a third-neighbor hopping term, while higher orders generate progressively longer odd-distance processes. We further presented a systematic Floquet construction that realizes long-range hopping arising from the q-deformation.

\par
The results, therefore, suggest that algebraic deformation of the kinetic operator provides a distinct route for modifying quasiperiodic localization and motivates further investigation of the associated spectral, critical, and multifractal properties of the resulting nonlocal quasiperiodic Hamiltonians. An interesting direction is to investigate the realization of the correlated hopping hierarchy generated by the \(q\)-deformation in synthetic quantum platforms. In particular, programmable long-range tunneling could provide a route to experimentally test the predicted energy-dependent localization structure. While realizing the complete \(q\)-dependent hierarchy remains a platform-specific engineering problem, the present results identify the localization phenomena that should arise once such correlated kinetic processes are implemented.

\section{Acknowledgement}
AG gratefully acknowledges Dr. Shaon Sahoo and Dr. Ranjan Modak for the useful discussions and previous collaborations on related topics. AG is also deeply grateful for the research facilities provided by IIT Tirupati. The author also thanks Pallabi Chatterjee for discussions on Space-fractional quantum mechanics.

\nocite{*}

\bibliography{manuscript}

\appendix
\section{Derivation of generalized Kinetic energy term}
\begin{figure*}
    \centering
    \includegraphics[width=0.95\linewidth]{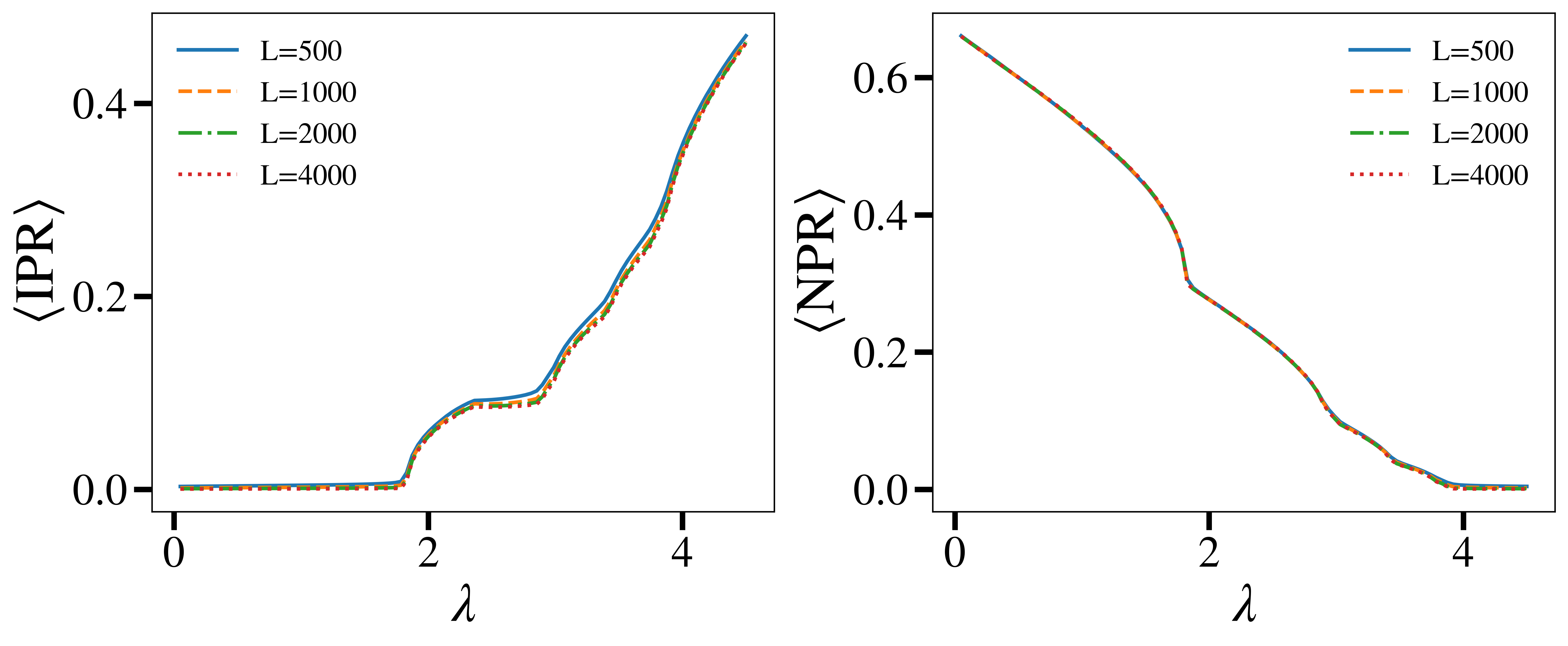}
    \caption{Finite-size convergence of localization diagnostics at $q=2.5$. Left: average inverse participation ratio $\langle \mathrm{IPR} \rangle$ versus $\lambda$. Right: average normalized participation ratio $\langle \mathrm{NPR} \rangle$ versus $\lambda$. Data are shown for four system sizes $N = 500, 1000, 2000$, and $4000$, each with a distinct linestyle. The close overlap of curves demonstrates excellent convergence and validates the reliability of the numerical results.}
    \label{fig:appendix_convergence}
\end{figure*}
The q-generalized number operator is defined as follows: 
\begin{equation}
    [n]_q =\frac{q^n -q^{-n}}{q-q^{-1}}.
\end{equation}
Replacing $\eta =\ln q$ we get the following expression:
\begin{equation}
    [n]_q =\frac{e^{n \eta}-e^{-n\eta}}{e^{\eta}-e^{-\eta}}.
\end{equation}
We can replace the number operator with any other clean operator($K$) to get the q-generalized operator and get the following expression:
\begin{equation}
    [K]_q=\frac{e^{K\eta}- e^{-K\eta}}{e^{\eta}- e^{-\eta}}
\end{equation}
\[[K]_q=\frac{\sinh{(K\eta)}}{\sinh{\eta}}\]

\section{Finite-Size Convergence of Localization Diagnostics}
To verify the robustness of our numerical results, we analyze the finite-size dependence of the average inverse participation ratio (IPR) and normalized participation ratio (NPR) as functions of the quasiperiodic potential strength $\lambda$ for a fixed deformation parameter $q=2.5$. Figure~\ref{fig:appendix_convergence} shows the corresponding plots for four system sizes $N = 500, 1000, 2000$, and $4000$.

Both quantities exhibit excellent convergence as the system size increases. The $\langle \mathrm{IPR} \rangle$ curves nearly collapse onto a single trajectory, with only minor deviations near the transition region around $\lambda \approx 2$. Similarly, the $\langle \mathrm{NPR} \rangle$ curves overlap almost perfectly across all $N$, confirming that finite-size effects are negligible. This consistency between IPR and NPR demonstrates that the localization diagnostics are well converged and accurately capture the thermodynamic-limit behavior of the q-deformed Aubry-Andr\'e model.

\end{document}